\documentclass{aa} 

\makeatletter
\usepackage{lastpage}
\renewcommand*\aa@pageof{, page \thepage{} of \pageref*{LastPage}}
\makeatother

\usepackage{amsmath}
\usepackage{epsf}
\usepackage{color}
\usepackage{enumitem}
\usepackage{mathtools}
\usepackage[mathcal]{eucal}
 \let\mathscr\relax

\usepackage{romannum}
\AtBeginDocument{\pagenumbering{arabic}}
\usepackage{graphicx}
\usepackage{txfonts}
\usepackage[colorlinks,citecolor=blue,hypertexnames=true,bookmarks=false,breaklinks=true]{hyperref}
\usepackage{tikz}
\usetikzlibrary{shapes.geometric, arrows}
\usepackage{multirow}
\usepackage{makecell}
\usepackage{tabularx}
\begin{document}

   \title{PRIMER: JWST/MIRI reveals the evolution of star-forming structures in galaxies at $z\leq2.5$}
   \titlerunning{JWST/MIRI mass--size relation}


   \author{Yipeng Lyu\inst{\ref{inst1}}
   \and Benjamin Magnelli\inst{\ref{inst1}}
   \and David Elbaz\inst{\ref{inst1}}
   \and Pablo G. P{\'e}rez-Gonz{\'a}lez\inst{\ref{inst2}}
   \and Camila Correa\inst{\ref{inst1}}
   \and Emanuele Daddi\inst{\ref{inst1}}
   \and Carlos G{\'o}mez-Guijarro\inst{\ref{inst1}}
   \and James S. Dunlop\inst{\ref{inst3}}
   \and Norman A. Grogin\inst{\ref{inst4}}
   \and Anton M. Koekemoer\inst{\ref{inst4}}
   \and Derek J. McLeod\inst{\ref{inst3}}
   \and Shiying Lu\inst{\ref{inst1}}
   }

   \institute{Universit{\'e} Paris-Saclay, Universit{\'e} Paris Cit{\'e}, CEA, CNRS, AIM, 91191, Gif-sur-Yvette, France\label{inst1}
   \and Centro de Astrobiolog{\'i}a (CAB), CSIC-INTA, Ctra. de Ajalvir km 4,
    Torrej{\'o}n de Ardoz, E-28850, Madrid, Spain\label{inst2}
    \and Institute for Astronomy, University of Edinburgh, Royal Observatory, Edinburgh EH9 3HJ, UK\label{inst3}
    \and Space Telescope Science Institute, 3700 San Martin Drive,
    Baltimore, MD 21218, USA\label{inst4}}

   \date{Received June 11th, 2024; accepted December 18th, 2024}

 
  \abstract
  {The stellar structures of star-forming galaxies (SFGs) undergo significant size growth during their mass assembly and must pass through a compaction phase as they evolve into quiescent galaxies (QGs). The mechanisms behind this structural evolution remain, however, poorly understood.
  }
  {We study the morphology of the star-forming components in SFGs to reveal the mechanisms that drive the structural evolution of their stellar components.
  }
    {We used high-resolution observations at 18 $\mu$m from the Mid-Infrared Instrument (MIRI) on board the \textit{James Webb} Space Telescope (JWST) taken as part of the Public Release IMaging for Extragalactic Research (PRIMER) survey to measure the morphology of star-forming components in 665 SFGs at $0<z<2.5$ and with $M_\ast\gtrsim 10^{9.5} M_\odot$. 
    We fit single S{\'e}rsic models to get the mid-infrared (MIR) structural parameters of these galaxies. The rest-frame optical morphology was taken from the literature and the effects of radial color gradients (due to dust or stellar aging) were corrected to obtain the intrinsic structural parameters for the stellar components of these galaxies.
    }
   {The stellar and star-forming components of most SFGs (66\%) have extended disk-like structures (S{\'e}rsic index, $n_\text{MIR}\sim0.7$ and $n_\text{optical}\sim1$; flat axis ratio distribution; hereafter called extended-extended galaxies) that are well aligned with each other and of the same size.
   Similar to the stellar components, the star-forming components of these galaxies follow a mass--size relation, with a slope of 0.12, and the normalization of this relation increases by $\sim0.23$ dex from $z\sim 2.5\text{ to }0.5$. At the highest masses ($M_\ast\gtrsim7\times10^{10} M_\odot$), the optical S{\'e}rsic index of these SFGs increases to $n_\text{optical}\sim2.5$, suggesting the presence of a dominant stellar bulge. Because their star-forming components remain in a disk-like structure, these bulges cannot have formed by secular in situ growth.
   We also observe a second population of galaxies lying below the MIR mass--size relation, with compact star-forming components embedded in extended stellar components. These galaxies are rare (15\%; called extended-compact galaxies) but become more dominant at high masses ($\sim30\%$ at $M_\ast>3\times10^{10} M_\odot$). The star-forming components of these galaxies are compact, concentrated ($n_\text{MIR}>1$), and slightly spheroidal ($b/a>0.5$), suggesting that this compaction phase can build dense stellar bulges in situ. 
   We identified a third population of galaxies with both compact stellar and star-forming components (19\%; called compact-compact galaxies). The density and structure of their stellar cores ($n_\text{optical}\sim1.5$; $b/a\sim0.8$) resemble those of QGs and are compatible with them being the descendants of extended-compact galaxies.} 
   {The structural evolution of the stellar components of SFGs is mainly dominated by an inside-out secular growth. However, this secular growth might be interrupted by compaction phases triggered by either internal or external mechanisms, which build dominant central stellar bulges as those of QGs.}
   {}


   \maketitle
%
 
\section{Introduction}
In recent years, significant progress has been made in our understanding of galaxy formation and evolution through multiwavelength extragalactic deep surveys. The current consensus is that galaxies in the Universe can be broadly classified into two main categories \citep[e.g., ][]{Ilbert2013}: blue star-forming galaxies (SFGs) and red quiescent galaxies (QGs). In general, the SFGs mainly reside along a tight correlation in the plane of the star formation rate (SFR) against stellar mass \citep[the so-called main sequence; e.g.,][]{Elbaz2007, Daddi2007, Noeske2007, Schreiber2015, Barro2019, Popesso2023}, suggesting a secular growth for these galaxies. In contrast, QGs show limited star formation activities; hence, they lie well below the main sequence. One other fundamental difference between SFGs and QGs lies in their morphology. The SFGs are mostly disks with an exponential light profile (i.e., with a S{\'e}rsic index, $n\sim1$), whereas local QGs are more like the elliptical galaxy. with a nearly de Vaucouleurs light profile \citep[i.e., $n\sim4$; ][]{Shen2003, Wuyts2011}. To unravel the origin of this morphological bimodality between SFGs and QGs, and to understand the quenching mechanisms connecting these two galaxy populations, it is imperative to track in detail their structural evolution over stellar mass and cosmic time. 

Earlier research studies have found that the rest-frame optical sizes of SFGs and QGs show very different stellar mass and redshift dependencies \citep{Shen2003, vdw2014}. The SFGs typically have a larger effective radius (i.e., a larger half-light radius) than QGs at low and intermediate stellar mass. However, SFGs follow a rest-frame optical mass--size relation with a much shallower slope than QGs. This results in very massive SFGs and QGs having similar sizes \citep{vdw2014}, although massive QGs usually exhibit higher central surface stellar mass density \citep{Barro2013, Barro2017}.
These morphological differences indicate that a significant phase of size compaction is required for SFGs to evolve into QGs, and both observations and simulations show that this compaction phase is often linked with the cessation of central star formation \citep{Dekel2006, Barro2013, Lang2014, Dekel2014, Tacchella2016, Barro2017, Lustig2021, Lapiner2023}. Nevertheless, the details within this process remain poorly understood, and at present there is lack of a well-defined galaxy structural transformation narrative.

According to spectral energy distribution (SED) models, the rest-frame optical ($\sim5000\AA$) size of galaxies can be used to trace their stellar emission, and thus provide a proxy for their stellar mass distribution, although corrections are still required due to radial color gradient effects \citep{Suess2022, vdw2023}. To predict the future distribution of these galaxies' stellar mass, it is essential, then, to examine the ongoing stellar mass buildup processes; that is, where star formation is occurring. Over the past decade, various methods have been employed to measure the star-forming components in galaxies. A direct approach is to utilize $\textit{H}\alpha$ emission radiated from galaxy $\textit{H}\Romannum{2}$ regions surrounding young, massive, and short-lived stars (e.g., OB stars). However, care must be taken here as, due to the obscuration of the rest-frame optical light by dust, one cannot simply equate sizes of the observed $\textit{H}\alpha$ emission with those of the intrinsic star-forming components within galaxies \citep{Pablo2003, Nelson2012, Nelson2016, Tacchella2015}, especially for massive galaxies that are more heavily dust-obscured \citep[e.g., ][]{Wuyts2011, Whitaker2012}. To correct for dust attenuation, the Balmer decrement ($\textit{H}\alpha$/$\textit{H}\beta$) or the UV slope ($\beta$) are often used. Multiple studies have measured such dust-corrected $\textit{H}\alpha$ emission and demonstrated that the star-forming components of low- and intermediate-mass galaxies (i.e., $\textit{M}_{\ast}\sim10^{9}-10^{10}M_\odot$) are slightly more extended than the corresponding stellar disks, favoring an inside-out growth scenario of SFGs \citep{Nelson2012, Tacchella2015, Dokkum2015, Wilman2020, Matharu2022, Shen2024}. In contrast, massive SFGs ($\textit{M}_{\ast}\geq10^{11}M_\odot$) are found to exhibit a centrally depressed star formation, which favors an inside-out quenching scenario for such galaxies \citep{Tacchella2018}.
While interesting, the size measurements of the star-forming components in these galaxies remain uncertain because of these complex dust attenuation corrections of $\textit{H}\alpha$ emission, which depend on the assumed geometry and composition of the dust particles. 

In lieu of relying on the complicated dust attenuation correction of $\textit{H}\alpha$ (or UV) emission, some studies have sought to measure the size of high-redshift star-forming components via their dust emission in the far-infrared (FIR) and submillimeter bands. They have observed these SFGs mainly with the Atacama Large millimeter/submillimeter Array (ALMA) at an observed-frame wavelength of around 870 $\mu$m \citep{Simpson2015,Hodge2016,Fujimoto2017,CG2018,Elbaz2018,Lang2019,Puglisi2019,Gullberg2019,Chang2020,Tadaki2020,Puglisi2021,CG2022}. Despite variations in the sample selection, these investigations come to the agreement that in massive dusty SFGs, the size of the dust-obscured star-forming component is generally smaller (by a factor of two or three) than that of the stellar component. The compact nature of the star formation process could be a driving factor in the evolution of such massive galaxies \citep{CG2022}. These results appear to be at odds with the ones obtained from the $\textit{H}\alpha$ emission line, probably due to differential dust attenuation across the galaxies \citep[with the cores being more dusty, see e.g.,][]{Nelson2016} or the disjointed sample selection procedures.
Indeed, it is essential to note that ALMA-detected objects are predominantly massive, high-redshift SFGs with stellar masses exceeding $10^{11}\textit{M}_{\odot}$ and redshifts greater than two. Consequently, for galaxies with low and intermediate stellar masses ($\textit{M}_{\ast}\sim10^{9}-10^{10}M_\odot$), the distribution of dust-obscured star-forming components remains to be explored.

To investigate the dust-obscured star formation distribution in SFGs of more typical stellar masses ($\textit{M}_{\ast}\sim10^{9}-10^{10}M_\odot$), an infrared instrument with high sensitivity, good angular resolution, and a large field of view is required. The Mid-Infrared Instrument \citep[MIRI;][]{Rieke2015, Wright2023} on board the \textit{James Webb} Space Telescope \citep[JWST;][]{Gardner2006} provides such an opportunity by conducting comprehensive deep sky surveys spanning from the observed-frame 5.6~$\mu$m up to 25.5~$\mu$m with unparalleled quality. This wavelength range covers the mid-infrared (MIR) region where the emission of large complex molecules (e.g., the polycyclic aromatic hydrocarbons; PAHs) dominates (also active galactic nuclei (AGNs) or compact starbursts). Previous studies have shown that for local SFGs, up to 20\% of the total infrared luminosity is emitted within MIR bands \citep{Smith2007}, and that the MIR emission of SFGs can serve as an accurate tracer for star formation in both the low- and high-redshift Universe \citep{Elbaz2011, Schreiber2018, Ronayne2023, Shivaei2024}. Compared to its predecessor, \textit{Spitzer}/MIPS \citep{Rieke2004}, MIRI offers up to ten times deeper sensitivity and much higher angular resolution (e.g., $0\farcs6$ at 18 $\mu$m for MIRI versus $6\farcs$0 at 24 $\mu$m for MIPS). JWST/MIRI can therefore efficiently identify galaxies with infrared luminosities lower than $10^{10}$ $L_\odot$ at redshifts up to 2 \citep{Kirkpatrick2023}. Utilizing MIRI observations from the Cosmic Evolution Early Release Science survey, \citep[CEERS; ][]{Finkelstein2023, Yang2023b}, \citet{Shen2023}, and \citet{Magnelli2023} demonstrated that JWST/MIRI can be used to measure the morphological structures of star-forming components for distant galaxies with stellar masses as low as $10^{9}M_\odot$ \citep[see also ][]{Liu2023}. They successfully measured the star-forming sizes of several dozens of SFGs in the redshift range $0.2<z<2.5$ using the single S{\'e}rsic model and compared these sizes with those of their rest-frame near-UV/optical stellar emission. 
Interestingly, \citet{Magnelli2023} found that when radial color gradients affecting rest-frame optical sizes are accounted for, the size of the stellar and dust-obscured star-forming components of SFGs are, on average, consistent with each other at all stellar masses. However, there is a small population of SFGs ($\sim15$\%) with a compact star-forming component embedded in a larger stellar structure. They suggest that this galaxy population could be the missing link between the common SFGs (with extended stellar and star-forming components) and the ones with compact stellar components \citep[often called blue nuggets;][]{Barro2013, Barro2017}. 

In this paper, we build upon the work conducted by \citet{Magnelli2023} and expand their measurements to a sample ten times larger, comprising about 665 sources compared to the previous 69. This extension has been made possible through the JWST Public Release IMaging for Extragalactic Research (PRIMER) survey \citep{Dunlop2021}. With this enlarged dataset of SFGs, we conducted a more comprehensive statistical analysis and delved into the redshift and mass dependencies of the star-forming sizes. To this end, we firstly modeled the MIR emission of these SFGs at $0<z<2.5$ with a single S{\'e}rsic model. Then, we compared these measurements with the characteristics of their stellar components traced by the rest-frame optical observation \citep{vdw2014}, corrected for color gradient effects \citep{Suess2019, vdw2023}. This comparative analysis enables us to retrieve information on how the stellar structures of these galaxies evolve with time and stellar mass, and provides insights into the nature of the galaxy size compaction and the star formation quenching.

This paper is structured as follows. In Sect. \ref{data}, we introduce our data, which includes the JWST-PRIMER MIRI sample and the corresponding optical counterparts from the CANDELS catalogs \citep{Grogin2011, Koekemoer2011}. This section also outlines the sample selection procedures, morphology measurements, and the SED fitting performed with CIGALE \citep{Boquien2019} to infer the physical properties of our galaxies. Section \ref{results} is dedicated to presenting our findings, ranging from simple optical and MIR morphological comparisons to the inference of the MIR mass--size relation and a comprehensive analysis of its redshift evolution. In Sect. \ref{discussion}, we compare our results with the ones found in the literature, discuss the robustness of our findings, and integrate our results into the broader context of galaxy evolution. Section \ref{summary} concludes the paper with a summary. 

Throughout this paper, we adopt a standard concordance $\Lambda$CDM cosmology, with $H_0=70$ km s$^{-1}$, $\Omega_\Lambda=0.7$, and $\Omega_M=0.3$. For all stellar mass and SFR estimations, we utilize a Chabrier initial mass function \citep[IMF;][]{Chabrier2003}. Stellar masses and SFRs taken from the literature adopting a different IMF were converted by multiplying them by a factor of 1/1.78 for a Salpeter IMF \citep{Salpeter1955} and a factor of 1/0.94 for a Kroupa IMF \citep{Kroupa2001}.

\section{Data and sample selection}
\label{data}
\subsection{PRIMER}
\label{PRIMER-MIRI}
\begin{figure}
   \centering
   \begin{minipage}[c]{0.96\columnwidth}
   \center{\includegraphics[width=0.6\linewidth]{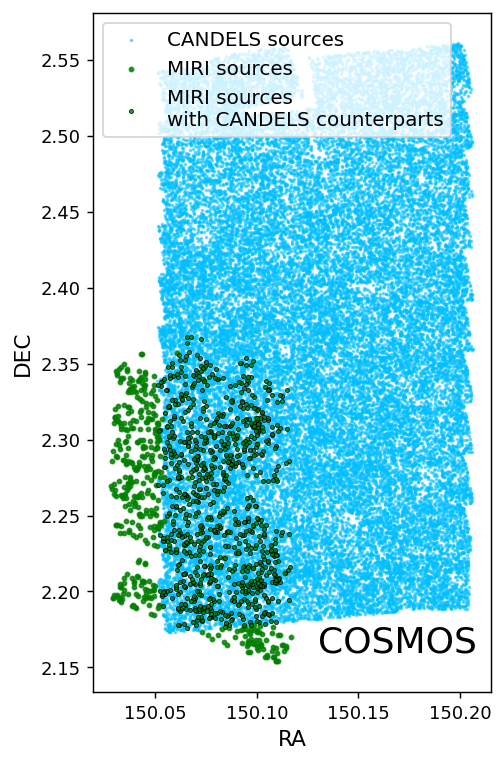}}
   \end{minipage}
   \begin{minipage}[c]{0.96\columnwidth}
  \center{\includegraphics[width=0.99\linewidth]{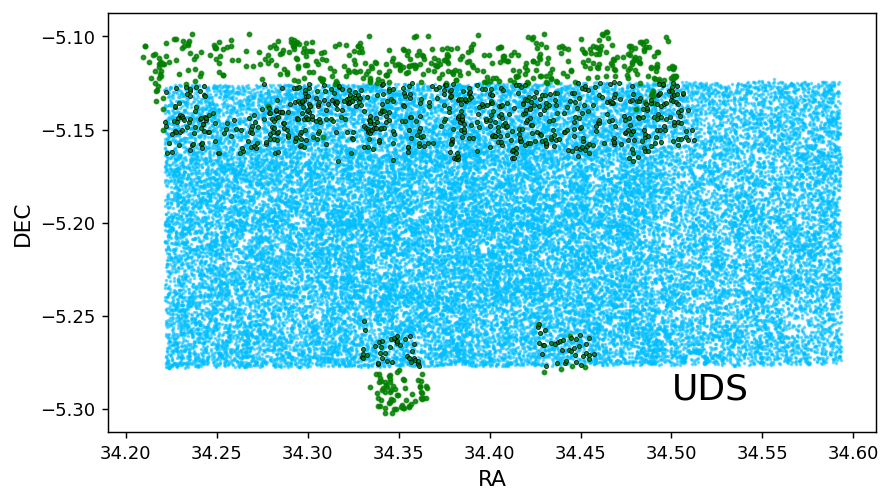}}
   \end{minipage}
\caption{Distribution of JWST PRIMER MIRI sources in both the CANDELS COSMOS (\textit{upper}) and UDS (\textit{lower}) deep fields. The green circles represent MIRI detection, corresponding to about half of the planned PRIMER MIRI coverage of COSMOS and UDS fields. Blue circles depict the CANDELS sources. Green circles with black edges are MIRI sources with CANDELS counterparts.}
\label{MIRI-CANDELS-maps}
\end{figure}

In this study, we leveraged data from the PRIMER survey \citep{Dunlop2021}, a public treasury program of JWST designed to provide an extensive and uniform coverage on parts of the COSMOS \citep{Scoville2007} and UDS \citep{Lawrence2007} deep fields, with both NIRCam and MIRI observations. The parts of the COSMOS and UDS fields targeted by PRIMER are those with deep WFC3 \textit{Hubble} Space Telescope (HST) coverage, from the Cosmic Assembly NIR Deep Extragalactic Legacy Survey \citep[CANDELS;][]{Grogin2011, Koekemoer2011}. 
The observing modes of PRIMER include ten bands in NIRCam (\textit{F090W}, \textit{F115W}, \textit{F150W}, \textit{F200W}, \textit{F277W}, \textit{F356W}, \textit{F444W}, and \textit{F410M}) and two bands in MIRI (\textit{F770W} and \textit{F1800W}). 
The observation, initiated in mid-November 2022, was mostly completed by August of 2023, with only a few pointings needing to be reobserved at the end of 2023 due to poor image quality from the first exposure. This paper specifically focuses on the MIRI maps obtained from PRIMER before May of 2023, covering approximately half of the planned COSMOS and UDS fields. The MIRI coverage of the CANDELS COSMOS and UDS fields adopted in this study (observations prior to May 2023) are shown in Fig. \ref{MIRI-CANDELS-maps}.

Details on the processing of the PRIMER MIRI observations can be found in \citet{Pablo2024a}. Here, we briefly describe the most important steps. 
The PRIMER MIRI imaging was reduced by the JWST rainbow pipeline developed within the European Consortium MIRI GTO Team, based on the JWST official pipeline. 
The background was homogenized using a so-called super-background strategy, and then was subtracted from the image. 
For the PRIMER data, we found that the official JWST pipeline did not manage to obtain an accurate world coordinate system (WCS) solution across the full mosaic (with systematic variations of $0\farcs2$). Therefore, we used the \textit{tweakreg} version provided by the CEERS collaboration \citep{Bagley2023} to align the individual frames before stacking and mosaicing. We then checked that WCS was consistent across all mosaics. 
At the end, we had our final science imagines, weighting maps, and RMS maps, with a pixel scale of $0\farcs06$, which corresponds to 0.48 kpc at the redshift of one. In this work, we mainly focus on the PRIMER-MIRI observations in the \textit{F1800W} band, which captures, for galaxies at $0<z<2.5$, the most prominent PAH features at the rest-frame 6.2, 7.7, 8.6, 11.3, and 12.7 $\mu$m \citep[e.g.,][]{Li2020}. In this band, the typical 5$\sigma$ sensitivity in an aperture of $0\farcs37$ is 23 mag AB.

The extraction and deblending of sources in our MIRI \textit{F1800W} maps were performed using the \texttt{python} package \texttt{photutils} \citep{photutils} with a smoothing Gaussian kernel. The connected pixels threshold was set to 20, meaning that a detected source has to contain at least 20 pixels. The connectivity was set to 8, implying that neighboring pixels can touch each other along either edges or corners. The deblend levels was set to 32 and the deblend contrast was set to 0.005 (i.e., deblending sources with a difference of at least 5.8 in magnitude).
The detection threshold was set to about two times the error in the map. This threshold was set after careful visual inspection to make sure that real-looking sources were not missed. We note that we also applied our MIRI \textit{F1800W} extraction maps to the photometry measurements of the PRIMER-MIRI \textit{F770W} observations after PSF-matching. These measurements were not used for the following morphological analysis, as this signal originates from both the stellar and the dust emission of galaxies, but they were used in our SED fit analysis (see Sect. \ref{AGN}), in which they add important extra information. 
In the end, our PRIMER MIRI \textit{F1800W} catalog contains 2517 sources (1254 objects in COSMOS and 1263 objects in UDS).
\subsection{Rest-frame optical counterparts}
To obtain the stellar mass, redshift, and rest-frame optical morphology of our MIRI-detected galaxies, we took advantage of the multiwavelength photometric data from CANDELS \citep{Grogin2011, Koekemoer2011}. CANDELS deep fields have been extensively observed by numerous ground and space-based multiwavelength surveys, spanning from hard X-rays to low-frequency radio, including but not limited to the observations from the Canada France Hawaii Telescope (CFHT)/MegaCam, Subaru/Suprime-cam, VLT/VISTA, VLT/HAWK-I, WFCAM/UKIRT, Mayall/NEWFIRM, \textit{Spitzer}/IRAC, \textit{Spitzer}/MIPS, \textit{Herschel}/PACS, and \textit{Herschel}/SPIRE, and most importantly the \textit{HST}/WFC3. The instruments employed on COSMOS and UDS vary slightly, leading to subtle difference in the associated multiwavelength catalogs \citep[for details, see ][]{Grogin2011, Koekemoer2011}.
For CANDELS COSMOS, we used the UV, optical, and NIR catalogs from \citet{Nayyeri2017} and far-IR data from \citet{Barro2019}. For CANDELS UDS, we adopted the UV, optical, and NIR catalogs from \citet{Galametz2013} and far-IR catalogs from \citet{Barro2019}. Redshift and stellar mass information for COSMOS were derived by \citet{Nayyeri2017}, while for the UDS this information was obtained from \citet{Santini2015}. In both cases, stellar mass and photometric redshift (when spectroscopic ones were unavailable) were obtained by modeling the multiwavelength photometry of each galaxy with various SED fitting codes. The typical error on these stellar masses is $\sim30\%$. By taking the median value outputs by these SED codes, the estimations are hence robust against the star formation history (SFH), assumed metallicity, extinction, and age parameterizations.
In terms of SFR, we chose to use the one from our fits with the SED fitting code CIGALE (see Sect. \ref{AGN}) because it makes the optimal and consistent use of all available panchromatic observations for each galaxy (including their MIR emission measured with MIRI). The typical error on our SFR is $\sim15\%$. 
As a sanity check, we cross-verified that the stellar masses taken from the literature were consistent with our own measurements obtained from CIGALE: the two estimations of stellar mass have a median ratio of 1.01 and a scatter of less than $0.05$ dex. 

We crossmatched our MIRI detected sample with the CANDELS catalogs based on their coordinates using \texttt{TOPCAT} \citep{Taylor2005}. The maximum error (searching radius) was set to $1\farcs0$. 
In Fig. \ref{MIRI-CANDELS-maps}, it is noted that the PRIMER MIRI observations do not fully overlap with the CANDELS deep fields. In fact, only 64\% of our PRIMER MIRI sources (1603) are in the sky area covered by CANDELS. Among these 1603 sources, 1591 have a counterpart in the CANDELS catalogs. 
This leaves us with a dozen MIRI sources in the CANDELS fields that do not have optical counterparts. After visual inspection, we find that four of them are completely undetected in the HST \textit{F160W} images but clearly detected at longer wavelengths, probably due to very heavy dust attenuation \citep{Wang2019}. 
The remaining ones are very faint in the HST \textit{F160W} images, and were hence missed by the CANDELS catalogs due to inevitable incompleteness at this signal level. The properties of these MIRI sources without CANDELS counterparts are thus fully consistent with the so-called optically dark or faint galaxies, whose redshifts are often higher than the one of the present study \citep[$z>3$;][]{Xiao2023}. For this reason, we simply ignored them in the rest of our analysis.

We took the rest-frame optical size of our galaxies from the catalogs of \citet{vdw2012}, who performed such morphological measurements for all objects detected by HST in the CANDELS deep fields. A single S{\'e}rsic model was assumed and fit for these objects using GALFIT \citep{Peng2002, Peng2010} in the closest rest-frame optical bands (\textit{F160W} for $z>1.5$ galaxies and \textit{F125W} for $z<1.5$ galaxies). These structural parameters include the S{\'e}rsic index ($n$), effective radius ($Re$, semimajor axis of the ellipse containing half of the total flux), axis ratio ($b/a$, semiminor axis over semimajor axis), and position angle ($PA$). To counteract the negative effect of size--color gradient correlating with the redshift and stellar mass of galaxies, we applied the morphological $K$-correction advocated in \citet{vdw2014}, which effectively converted all measured sizes at a given observed frame wavelength into rest-frame 5000 $\AA$ sizes,
\begin{equation}
Re_\text{opt}=Re_{\text{F}}\times\left(\frac{1+z}{1+z_p}\right)^\alpha,
\end{equation}
\text{with}
\begin{equation}
\alpha=-0.35+\text{0.12 }z-0.25\text{ log}(\textit{M}_{\ast}/{10^{10}\textit{M}_{\odot}}),
\end{equation}
where $Re_{\text{F}}$ is the size measured in the \textit{F125W} band when $z<1.5$ and in the \textit{F160W} band when $z>1.5$, and $z_p$ is the so-called pivot redshift (1.5 for \textit{F125W} and 2.2 for \textit{F160W}). As for the $\alpha$ above, any positive value was set to zero. We note that these corrections are minor for high-redshift, low-mass galaxies (less than 0.1 dex), but for low-redshift and high-mass galaxies the corrections can reach up to 0.15 dex.

Finally, we had to take into account the fact that the half-light radius at rest-frame $5000\AA$ were not yet perfect proxy for the half-mass radius of SFGs \citep{Suess2019, vdw2023}, due to negative radial dust attenuation gradients (i.e., the central galaxy region is more dust-attenuated than its outskirts) or stellar age gradients (the central galaxy region is predominantly occupied by older stellar populations than its outskirts). To this end, we simply used the empirical mass-dependent corrections from \citet{vdw2023}, which were deduced by comparing the optical half-light radius with the half-mass radius of $\sim500$ SFGs at $0.5<z<2.3$ derived from multiwavelength light profiles from HST and JWST images. These corrections imply that the half-mass radius of SFGs with a stellar mass of $10^{10}M_\odot$ and $10^{11}M_\odot$ are smaller by $\sim10\%$ and $\sim40\%$ than their rest-frame optical half-light radius, respectively. We note that the stellar mass estimations from \citet{vdw2023} were obtained with Prospector using a nonparametric SFH. This type of SFH is known to lead to slightly higher stellar mass compared to the delayed-$\tau$ SFH used in our analysis \citep[$\sim0.1$dex][]{Ciesla2023}. However, the half-light to half-mass corrections from \citet{Suess2022} obtained using a delayed-$\tau$ SFH too do not suggest that there is a need to include such stellar mass rescaling (i.e., +0.1 dex) before using \citet{vdw2023} corrections. Therefore, we adopted the original \citet{vdw2023} corrections and simply added in quadrature extra 10\% uncertainties to our optical corrected size measurements.

\subsection{Active galactic nucleus identification}
\label{AGN}
\renewcommand{\arraystretch}{1.6}
\begin{table*}
\centering
\caption{CIGALE model parameters}
\label{SEDsettings}
\begin{tabular}{l l l}        
\hline\hline                 
Module & Parameter & Values \\     
\hline                        
   \multirow{2}*{\makecell[l]{Star formation history \\ \textit{sfhdelayed}}} & Stellar e-folding time $\tau$ & 0.5, 1, 5 Gyr \\     
     & Stellar age $t$ & 1, 3, 5, 7 Gyr\\
     \hline 
   \multirow{2}*{\makecell[l]{Simple stellar population \\ \textit{\citet{Bruzual2003}}}} & Initial mass function & \citet{Chabrier2003}\\
    & Metallicity $Z_{\odot}$& 0.02\\
    \hline 
   \multirow{3}*{\makecell[l]{Dust attenuation \\ \textit{\citet{Calzetti2001}} }}& $E(B-V)_\text{line}$ & \makecell[l]{0, 0.05, 0.1, 0.15, 0.2, 0.25, 0.3, 0.35 \\ 0.4, 0.45, 0.5, 0.55, 0.6, 0.65, 0.7, 0.75, 0.80}\\
    & $E(B-V)_\text{Factor}$ & 1\\ 
    \hline 
   \multirow{4}*{\makecell[l]{Dust emission \\ \textit{\citet{Draine2007}} }}& PAH mass fraction $q_{\text{PAH}}$& 0.47, 1.12, 1.77, 2.50, 3.90\\
    & Minimum radiation field $U_\text{min}$ & 5, 12, 20, 30, 40\\ 
    & Power-law slope $dU/dM \propto U^\alpha$ & 2\\ 
    & Fraction of PDR emission & 0.01, 0.1, 0.5, 0.99\\
    \hline 
   \multirow{3}*{\makecell[l]{AGN emission \\ \textit{\citet{Stalevski2016}} }}& Viewing angle & 30, 70\\
    & AGN fraction $f_\text{AGN}$ &\makecell[l]{0,0.01,0.03,0.05,0.1,0.2,0.3, \\ 0.4, 0.45, 0.4,0.5,0.6,0.7,0.8,0.9} \\ 
    & Wavelength range $\lambda_\text{AGN}$ & 3-30 $\mu$m\\ 
\hline                                   
\end{tabular}
\end{table*}
\renewcommand{\arraystretch}{1}
The emission of AGNs can dominate the optical and/or MIR emission of SFGs \citep{Koratkar1999, Lyu2022, Yang2021}, and thus significantly bias the structural parameters inferred at these wavelengths. To ensure that the size measured in the MIR band indeed originates from star-forming components, it is imperative to identify galaxies with significant AGN emission in this band and eliminate them from our sample. To this end, we fit the UV-to-FIR photometry of each galaxy in our MIRI$\times$CANDELS sample using the SED fitting code CIGALE \citep{Boquien2019}. To set the various modules and parameters of CIGALE, we followed the description in \citet{Magnelli2023} and \citet{Yang2023}, but with a much finer $E(B-V)$ and AGN grid (see Table \ref{SEDsettings}). In brief, we fixed the redshift according to the CANDELS catalogs and used the standard delayed-$\tau$ SFH.
We used the BC03 simple stellar population \citep[SSP;][]{Bruzual2003}, assuming a \citet{Chabrier2003} IMF with solar metallicity ($Z_\odot=0.02$). We utilized the \citet{Calzetti2001} attenuation law 
and the dust emission model of \citet{Draine2007}, assuming the energy balance.
For the AGN emission, we used the clumpy torus model from \citet{Stalevski2016}, considering both type 1 and type 2 AGNs corresponding to viewing angles of 30 and 70 degree, respectively, though some studies suggest that including type 1 AGNs in the CIGALE fitting causes significant model degeneracy versus stellar emission in the UV/optical \citep{Ni2021}. The ratio between AGN luminosity and total luminosity was parameterized by the fraction of AGN emission ($f_{\text{AGN}}$), which was calculated in the wavelength range between 3 and 30~$\mu$m.

In addition to these CIGALE SED fits, we also used the X-ray luminosity of our galaxies to identify any potential AGN-dominated systems. To do this, we used the Chandra X-ray legacy survey \citep{Civano2016, Marchesi2016} for COSMOS field and X-UDS Chandra observations \citep{Kocevski2018} for UDS field. After cross-matching, there are 29 sources that have X-ray detections, 27 of which have an X-ray luminosity larger than $10^{42.5}$ $\text{erg s}^{-1}$, which is usually treated as the threshold between star-forming dominated galaxies and AGN-dominated galaxies \citep[see in, e.g.,][]{Yang2018}. Among these X-ray AGNs, 13 of them have $f_{\text{AGN}}\leq0.1$, and 14 of them have $f_{\text{AGN}}>0.1$. Among the latter 14 galaxies, 11 of them are all classified by CIGALE as type-2 AGNs, whose luminosity is significantly obscured by the dust torus and that thus strongly re-emit in the MIR band.

Combining both SED fitting results and X-ray luminosity information, we classified a galaxy as a SFG only if it has $f_{\text{AGN}}\leq0.1$ and its X-ray luminosity is smaller than $10^{42.5}$ $L_\odot$ \citep[as, e.g., in ][]{Yang2023}. We identified 237 AGNs and were left with 1354 pure SFGs for our subsequent MIR structural analysis. Very few AGNs (less than 10\%) were identified only from their X-ray luminosity. The AGN incidence in our sample is consistent with what is found in \citet{Magnelli2023} but slightly lower than in \citet{Yang2023}, who found an AGN incidence of 142 out of 560 MIRI detected galaxies. The slight disagreement (16\% vs. 25\%) may be explained by the fact that  \citet{Yang2023} have taken into account six MIRI bands covering from \textit{F770W} to \textit{F2100W} in their analysis, which better constrains the AGN emission fraction than using only \textit{F770W} and \textit{F1800W}, as is also proved in the JWST/MIRI simulation of \citet{Yang2021}. In Fig. \ref{galfit-fitting}, we present three examples of SED fitting from the AGN-dominated galaxies to the pure SFGs. It is noteworthy that for galaxies dominated by AGN emission, the typical PAH features between 3 to 20 $\mu$m are completely concealed by the underlying the AGN continuum.

\subsection{MIR morphology measurements}
\label{galfit}
\subsubsection{Point spread function}
As in \citet{Shen2023} and \citet{Magnelli2023}, in this work we performed our structural analysis in the MIR using the theoretical MIRI point spread function (PSF) produced by the \texttt{WebbPSF} model \citep{Perrin2014} built in \texttt{python}. However, unlike these previous works, which were based on the CEERS MIRI map with small sky coverage ($\sim8$ arcmin$^{2}$), the relatively wide area covered by our PRIMER MIRI maps allows for the presence of a few ideal point sources (i.e., stars, quasars) that can be used to create an effective PSF. With this effective PSF, we can test at first order the accuracy of the theoretical model from \texttt{WebbPSF}.
Firstly, we preselected high signal-to-noise ratio (S/N $>100$) sources in the MIRI images with the \texttt{photutils} task \texttt{Find\_peak}. Then, we visually selected the clean point sources among them, excluding, for example, some of bright stars with diffraction patterns strongly contaminated by neighboring sources or bad pixels. Finally, we created the final effective PSFs by aligning and stacking the three and two perfect point sources selected in COSMOS and UDS, respectively. These effective PSFs were constructed with a pixel scale of $0\farcs06$, using the \texttt{python} package \texttt{photutils} and its task \texttt{EPSFBuilder} \citep{photutils}.
In the end, we find that the effective PSFs and those modeled from \texttt{WebbPSF} are consistent within 5\% in terms of their full width at half maximum (FWHM). Their major and minor axis and position angle are also consistent and well aligned. This finding is consistent with the results of \citet{Libralato2024}, who constructed MIRI effective PSFs using various JWST Cycle-1 public surveys and found a good agreement between effective PSFs and \texttt{WebbPSF} models at 18~$\mu$m.
However, the small number of ideal point sources found in our fields (two and three) results in the fact that slight changes in the input parameters of the \texttt{EPSFBuilder} task (e.g., the oversampling rate and the number of iterations) can have impacts on the output effective PSF FWHM.
As we are working in the MIR domain, where the angular resolution is considerably lower than in the near-infrared (NIR) or optical, some galaxies have sizes very close to these angular resolution limits. This implies that for these SFGs, even very slight deviations (one pixel; $\sim10\%$) in the FWHM can translate into noticeable changes in their size measurements. 
We find that these deviations in the FWHM of our different effective PSF realizations are at almost the same level as the differences between the effective PSFs and those from \texttt{WebbPSF}. 
Therefore, to enable our results to be more easily reproduced, we decided to use PSFs generated by \texttt{WebbPSF} for our subsequent MIRI morphological analysis. We compared our main results with those obtained using the effective PSFs generated by EPSFBuilder and found that the choice of PSFs had little impact on our conclusions. Nevertheless, we added in quadrature an additional 20\% uncertainty to our size measurements, typical of the dispersion in size measurements obtained using these different effective PSFs. Finally, we note that the FWHM of WebbPSF at \textit{F1800W} is $0\farcs6$. This corresponds to an effective radius of about $0\farcs2$ (assuming a Gaussian profile), which is smaller than the typical size in our sample ($\sim0\farcs3$).

\subsubsection{GALFIT fitting} 
We utilized GALFIT \citep{Peng2002,Peng2010} to fit the MIR emission of our pure SFGs sample, with a 2D single-component S{\'e}rsic model, as was done in \citet{Shen2023}. In contrast, \citet{Magnelli2023} used a Markov chain Monte Carlo (MCMC) approach to fit this 2D S{\'e}rsic model, which gives a more reasonable estimation of the uncertainties of the fitting parameters. However, MCMC is very computationally expensive, and thus not suitable for our large sample. We compared the performance between GALFIT and MCMC using simulations, and found that these two methods exhibit consistent results. We also note that although GALFIT would allow us to fit a more complex two-component (e.g., bulge+disk) S{\'e}rsic model to the MIR emission, the angular resolution and S/N of these MIRI detections would limit such an analysis to the few brightest and most extended SFGs in our sample, hampering the representativeness of our results. We defer this two-component analysis to future work.

For the GALFIT fitting, the initial parameters (positions, magnitude, size, etc.) of each galaxy were set based on the segmentation map generated by \texttt{photutils} during the source extraction (see Sect. \ref{PRIMER-MIRI}).
To balance between speeding up our fits, mitigating contamination from surrounding bright sources, and dealing with source deblending issues, we chose to fit the central target galaxy and simultaneously all galaxies in an image cutout with a size of up to $200\times200$ pixels ($12^{\prime\prime}\times12^{\prime\prime}$), the specific choice depending on the area of the target galaxy estimated by \texttt{photutils}. At the end, we only kept the fitting results of the central target.
To perform this fitting analysis, we provided GALFIT with the MIRI error map generated by the JWST pipeline. Letting GALFIT generate its own noise estimations was unfortunately impractical due to the lack of necessary information \citep{vdw2014}.
When GALFIT fails, typically indicated by a ``warning'' message,
we perturbed the initial parameters, by changing the cutout size or using a smaller initial magnitude estimation, and attempted to refit for at most ten times. For sources still failing after ten runs, we stopped the process and flagged them as ``non-fitting'' sources.

Sources successfully fit by GALFIT have each of their output parameters labeled as ``doubtful'' (marked by GALFIT with a ``$\ast$'' sign) or ``robust'' (no ``$\ast$'' sign). Following \citet{vdw2014}, we used this information to flag the GALFIT structural parameters of a given galaxy into ``good'' (all seven parameters are robust), ``acceptable'' (the S{\'e}rsic index and effective radius are robust, while some of other parameters are doubtful, which is quite rare in our sample), or ``bad'' (one of the S{\'e}rsic index or the effective radius is doubtful). We also flagged a source as bad once one of its fitting parameters reached the setting boundary values (e.g., with a S{\'e}rsic index of 0.2 or 8, or an ellipticity of 0.1). 
In addition, by visual inspection, we found that five galaxies in the sample have blending issues in MIR due to the worse resolution of the JWST \textit{F1800W} image. These blending issues correspond to several close-by optically detected galaxies that look like one single source in the MIR. We also flagged these problematic sources as bad.

\begin{figure*}
\centering
\text{Examples of the morphological fitting and SED fitting}\par\medskip
    \begin{minipage}[c]{0.75\columnwidth}
        \centering
        \center{\includegraphics[width=0.9\linewidth]{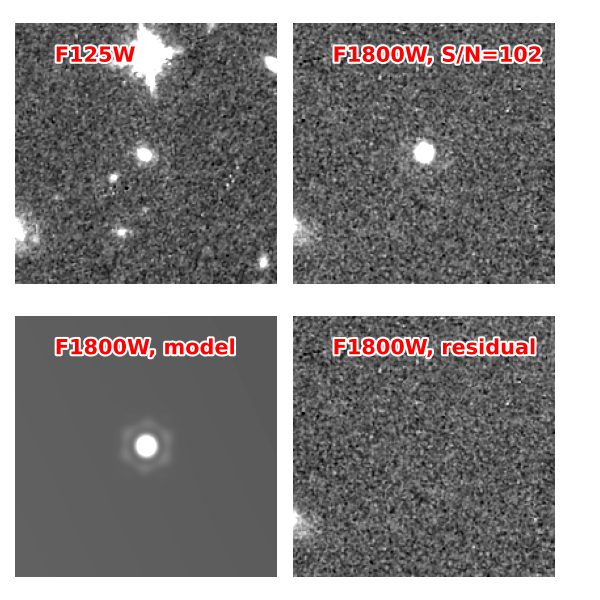}}
    \end{minipage}
        \begin{minipage}[c]{1.14\columnwidth}
        \centering
    \center{\includegraphics[width=0.95\linewidth]{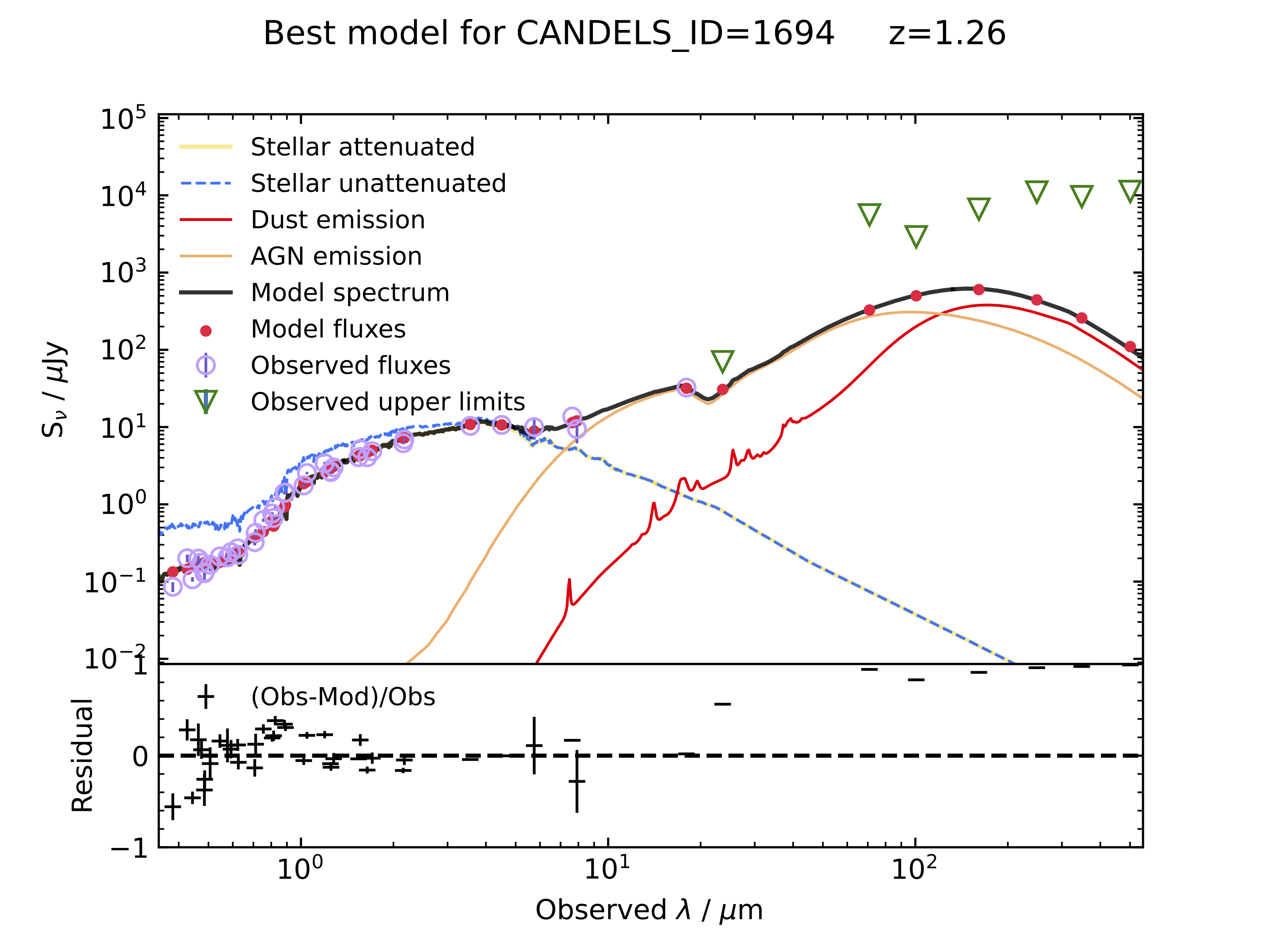}}
    \end{minipage}
    \begin{minipage}[c]{0.75\columnwidth}
        \centering
        \center{\includegraphics[width=0.9\linewidth]{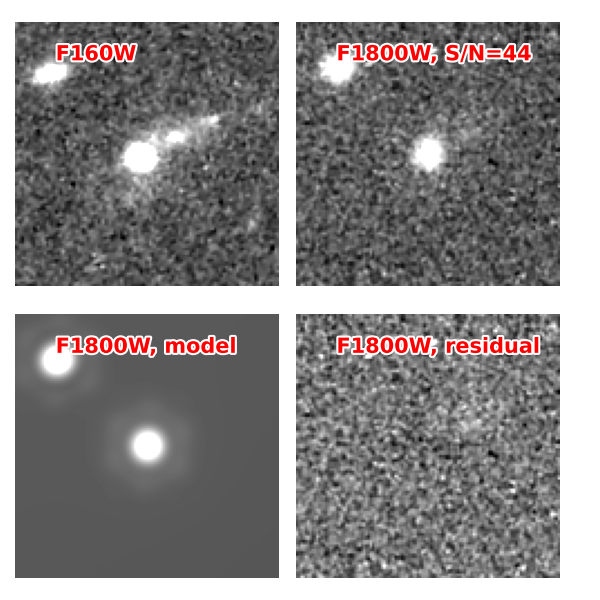}}
    \end{minipage}
    \begin{minipage}[c]{1.14\columnwidth}
    \center{\includegraphics[width=0.95\linewidth]{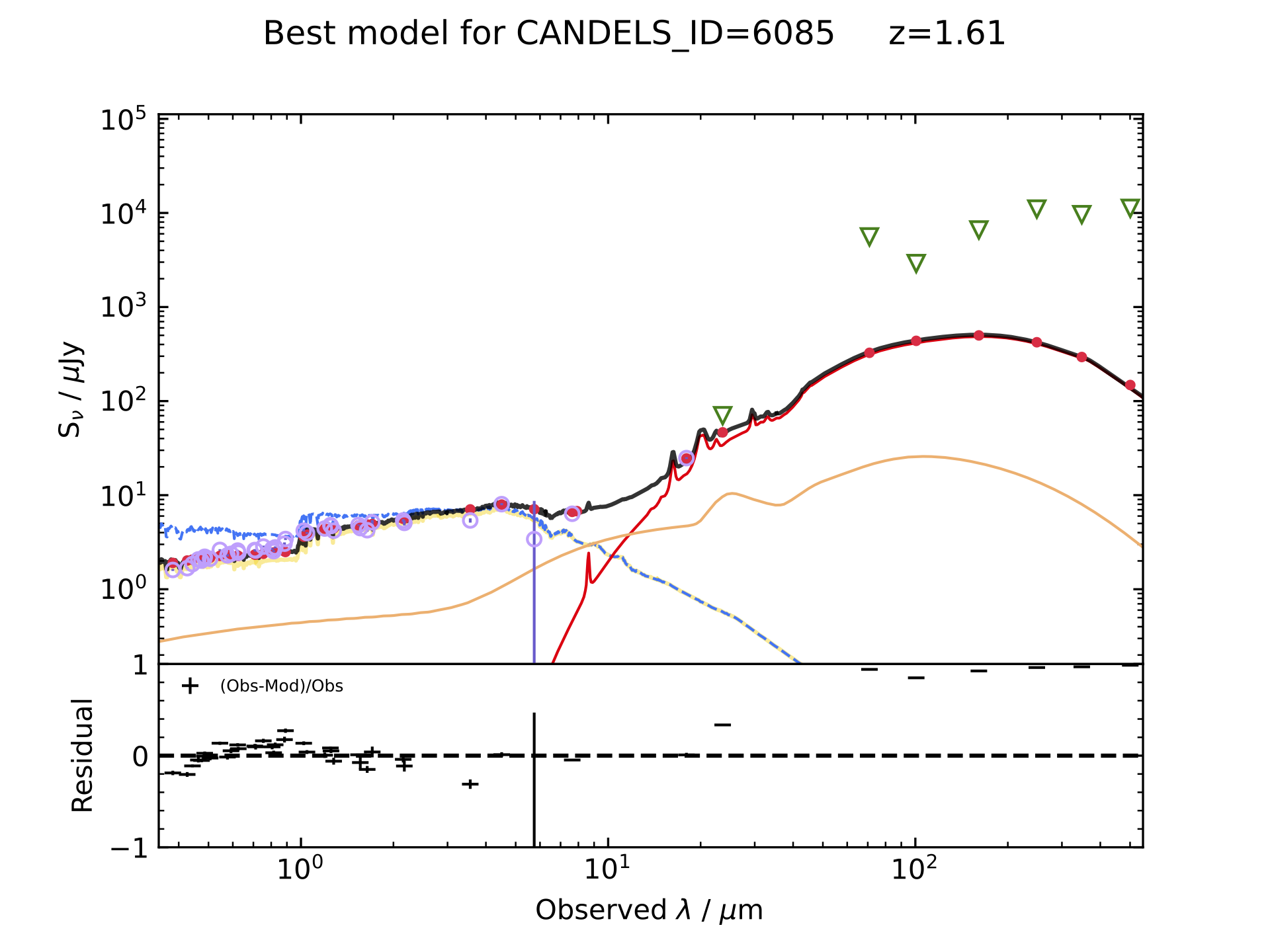}}
    \end{minipage}
    \begin{minipage}[c]{0.75\columnwidth}
        \centering
        \center{\includegraphics[width=0.9\linewidth]{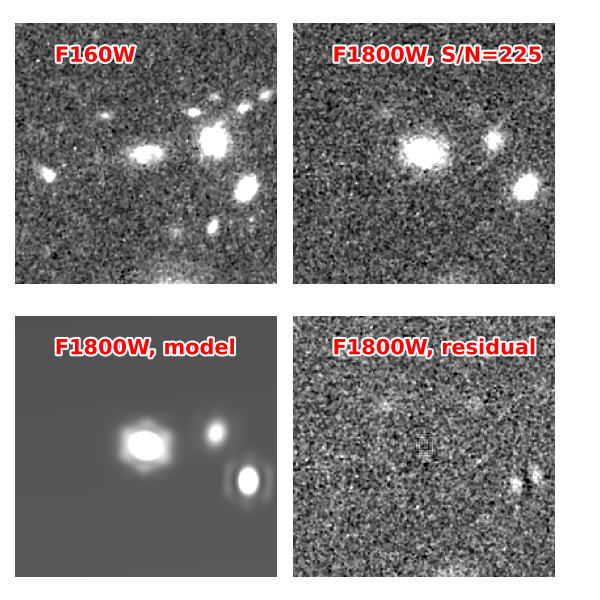}}
    \end{minipage}
    \begin{minipage}[c]{1.14\columnwidth}
        \centering
    \center{\includegraphics[width=0.95\linewidth]{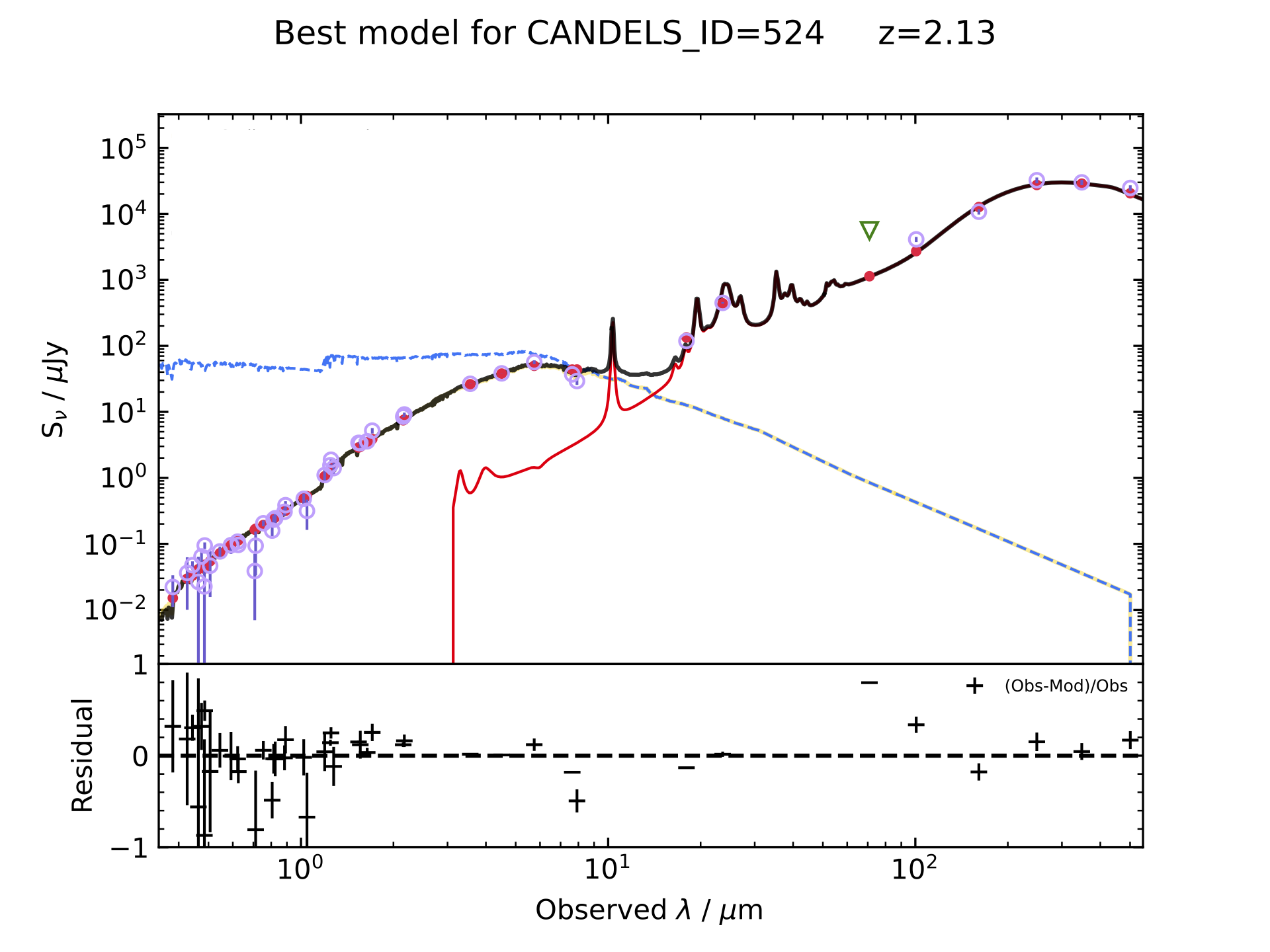}}
    \end{minipage}
\caption{Left: $10^{\prime\prime}\times10^{\prime\prime}$ cutouts of the HST \textit{F125W} ($z<1.5$) or \textit{F160W} ($z>1.5$) and MIRI \textit{F1800W} images (\textit{upper panels}), and of the GALFIT model and residual images (\textit{lower panels}), for three galaxies in our MIRI$\times$CANDELS sample (COSMOS field). Right: Best SED fitting of these galaxies as obtained by CIGALE \citep{Boquien2019}. Blue and red curves are stellar and dust emission; orange curves indicate the emission coming from the AGN component. From top to bottom, the AGN fraction is decreasing, from 70\%, to 10\%, and then 0\%. For galaxies dominated by AGNs, typical PAH features between 3 to 20 $\mu$m are hidden by the AGN continuum.}
\label{galfit-fitting}
\end{figure*}

We find that, except for objects with blending issues, the rest of the bad and non-fitting objects are sources with very small S/N ($<10$) or sources with missed inner pixels. We excluded all these bad and non-fitting galaxies from our subsequent analysis (see Sect. \ref{Final sample} for the impact of these exclusions on the representativity of our final sample).
As in \citet{vdw2014}, by visual inspection, we find that acceptable objects are not necessarily associated with terrible fits. Furthermore, we verified that the magnitudes produced by GALFIT and those from \texttt{photutils} for these acceptable galaxies were consistent with each other within $3\sigma$ (this also being true for galaxies flagged as good). We therefore decided to use all the galaxies flagged as good and acceptable in our subsequent MIRI morphological analysis. Our final S{\'e}rsic model contains seven parameters: positions ($x$, $y$), magnitude ($mag$), $Re$, $n$, $b/a$, and $PA$. For illustration, in the left of Fig. \ref{galfit-fitting}, we show examples of GALFIT MIR fits for three galaxies (from AGN-dominated to purely star-forming), along with their rest-frame optical images. It is noted that our fitting strategy with GALFIT provides accurate fits even in a crowded environment. 

\subsubsection{GALFIT simulation}
\label{galfitsimulation}
Regardless of the fitting methods, obtaining accurate structural parameters for faint sources is challenging, and GALFIT is no exception. In \citet{vdw2014}, GALFIT was applied on galaxies that are at least two magnitudes brighter than the $5\sigma$ limit (S/N $\sim30$). A similar limit (S/N $\sim40$) was used in \citet{Shen2023}. To explore the S/N thresholds above which all seven parameters of GALFIT S{\'e}rsic model fitting can be trusted, we conducted a set of Monte Carlo simulations with 10000 mock galaxies. The structures of MIR emission of our simulated galaxies were taken from the catalogs of \citet[][assuming MIR and rest-frame optical galaxies have similar morphology distribution]{vdw2014}, to which we added small random perturbations (20\% variations). To create these mock MIR emission profiles, we used the \texttt{python} packages \texttt{Sersic2D} and \texttt{petrofit}, and adopted an over-sampling rate of ten. These simulated galaxies were then convolved with the PSF generated by \texttt{WebbPSF} and projected into random empty regions of the original PRIMER-MIRI maps. These empty regions were identified based on the segmentation map generated by \texttt{photutils}.
Finally, we fit these simulated galaxies with GALFIT following two different strategies: 
\begin{enumerate}
    \item Fitting the images with all seven parameters of the single 2D S{\'e}rsic model (hereafter referred to as strategy 1).
    \item Fixing the S{\'e}rsic index to the median value of the group (in the rest-frame optical it is one, while in MIR it is $0.7$; see Sect. \ref{Sersic index}) and fitting the remaining six structural parameters (hereafter referred to as strategy 2).
\end{enumerate} 

\begin{figure*}
   \centering
   \begin{minipage}[c]{0.8\columnwidth}
        \centering
        \center{\includegraphics[width=0.99\linewidth]{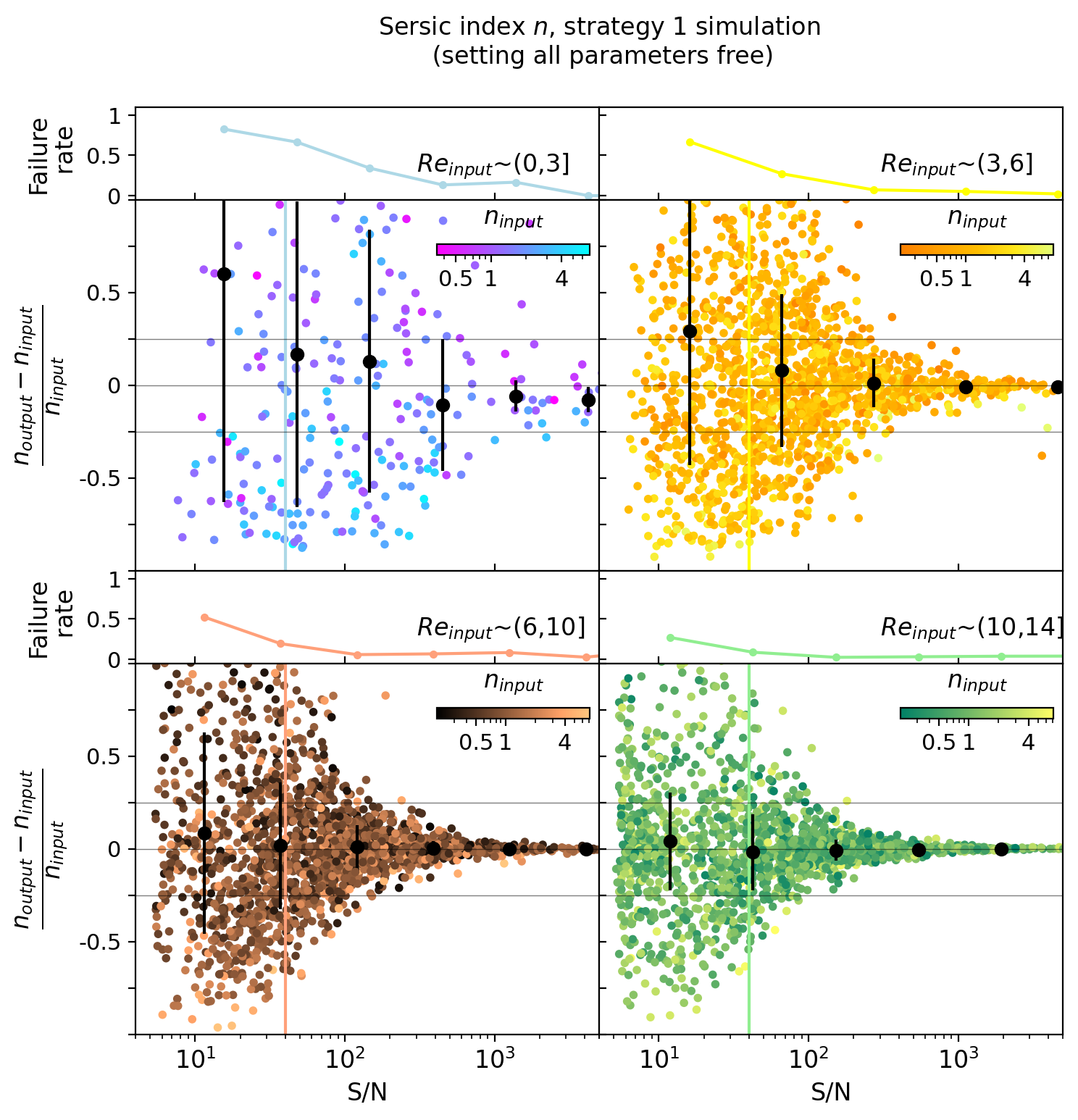}}
   \end{minipage}
   \begin{minipage}[c]{0.8\columnwidth}
        \centering
        \center{\includegraphics[width=0.99\linewidth]{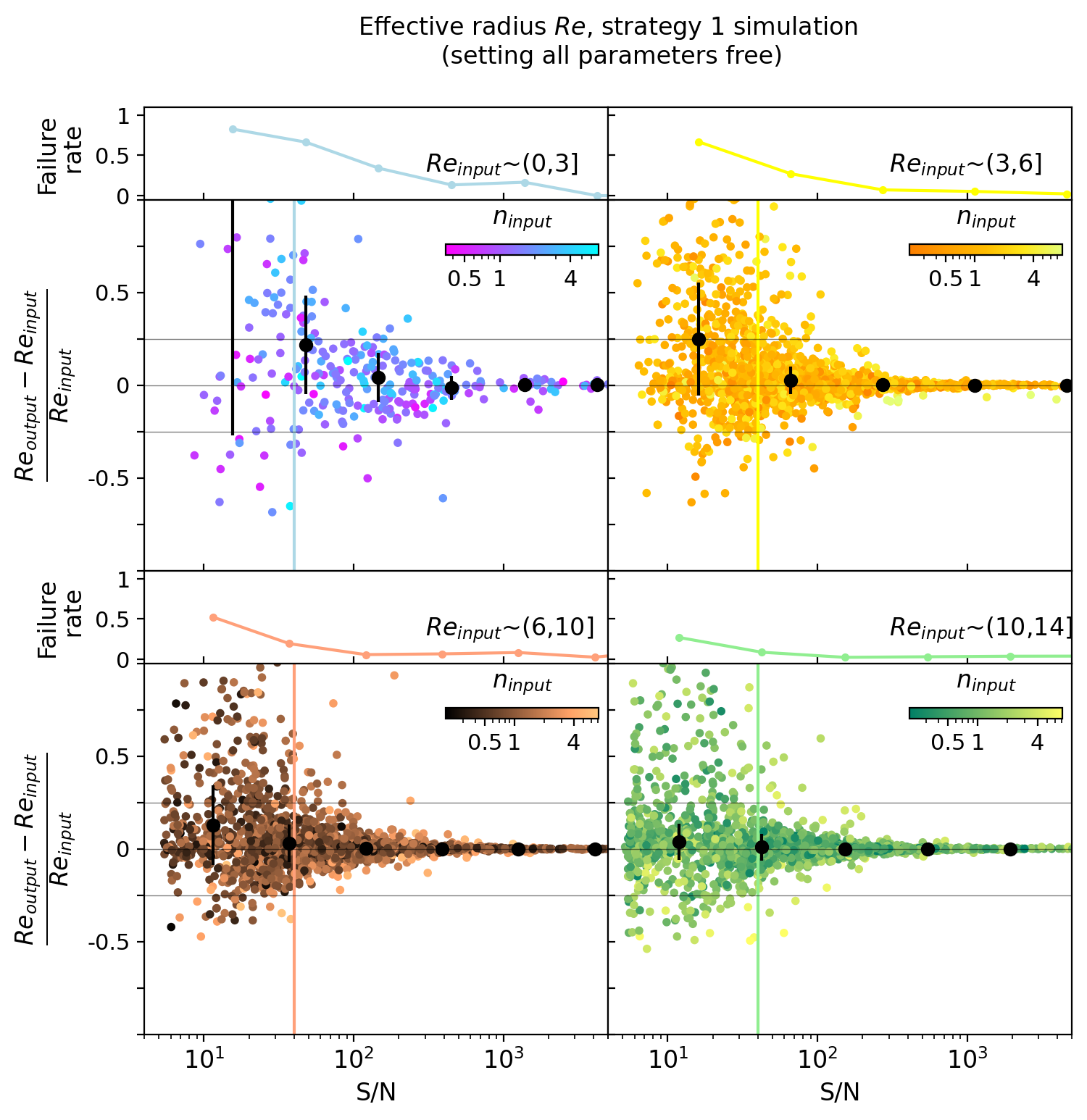}}
   \end{minipage}
   \begin{minipage}[c]{0.8\columnwidth}
        \centering
        \center{\includegraphics[width=0.99\linewidth]{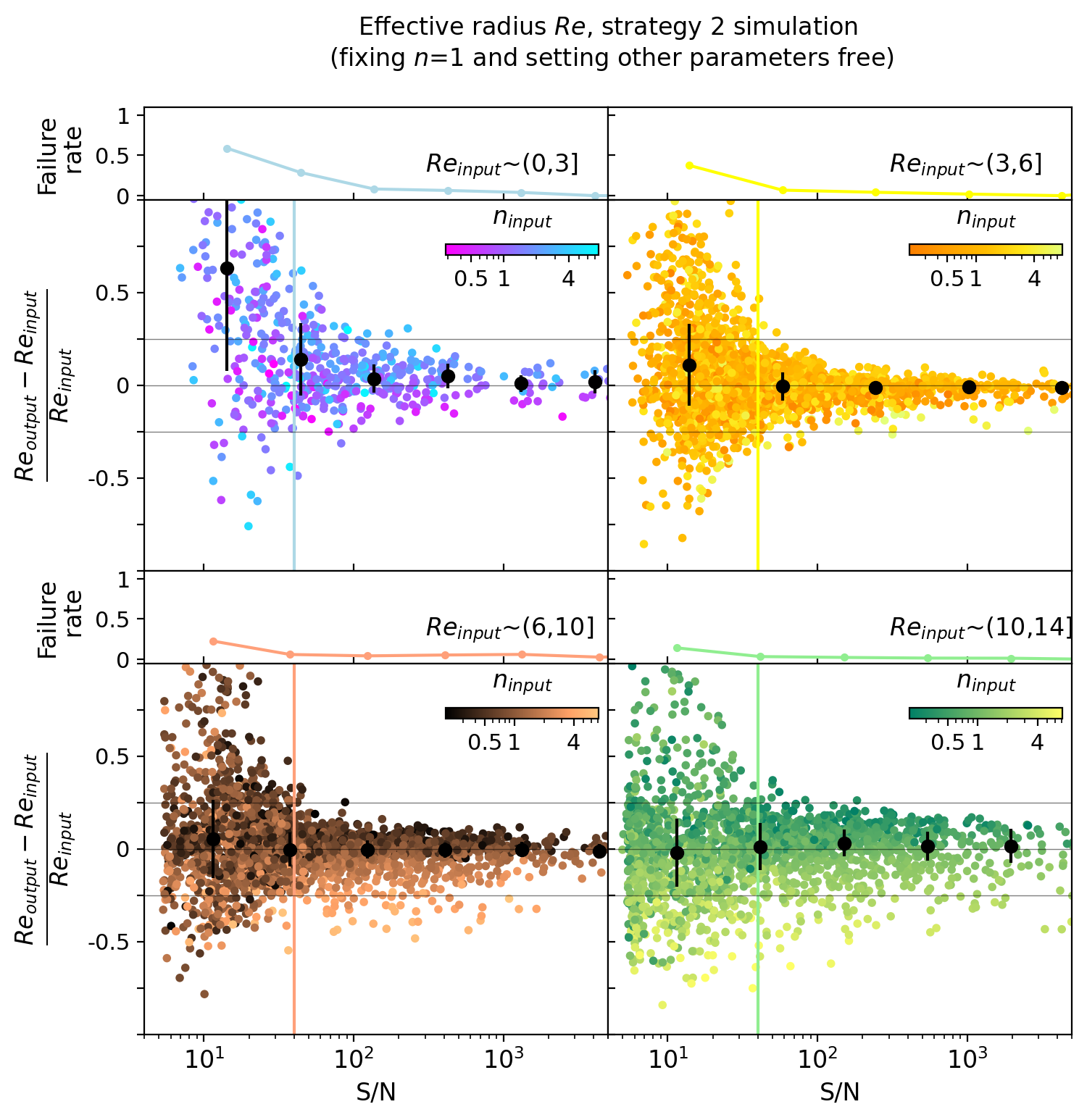}}
   \end{minipage}
   \begin{minipage}[c]{0.8\columnwidth}
        \centering
        \center{\includegraphics[width=0.98\linewidth, height=1.02\linewidth]{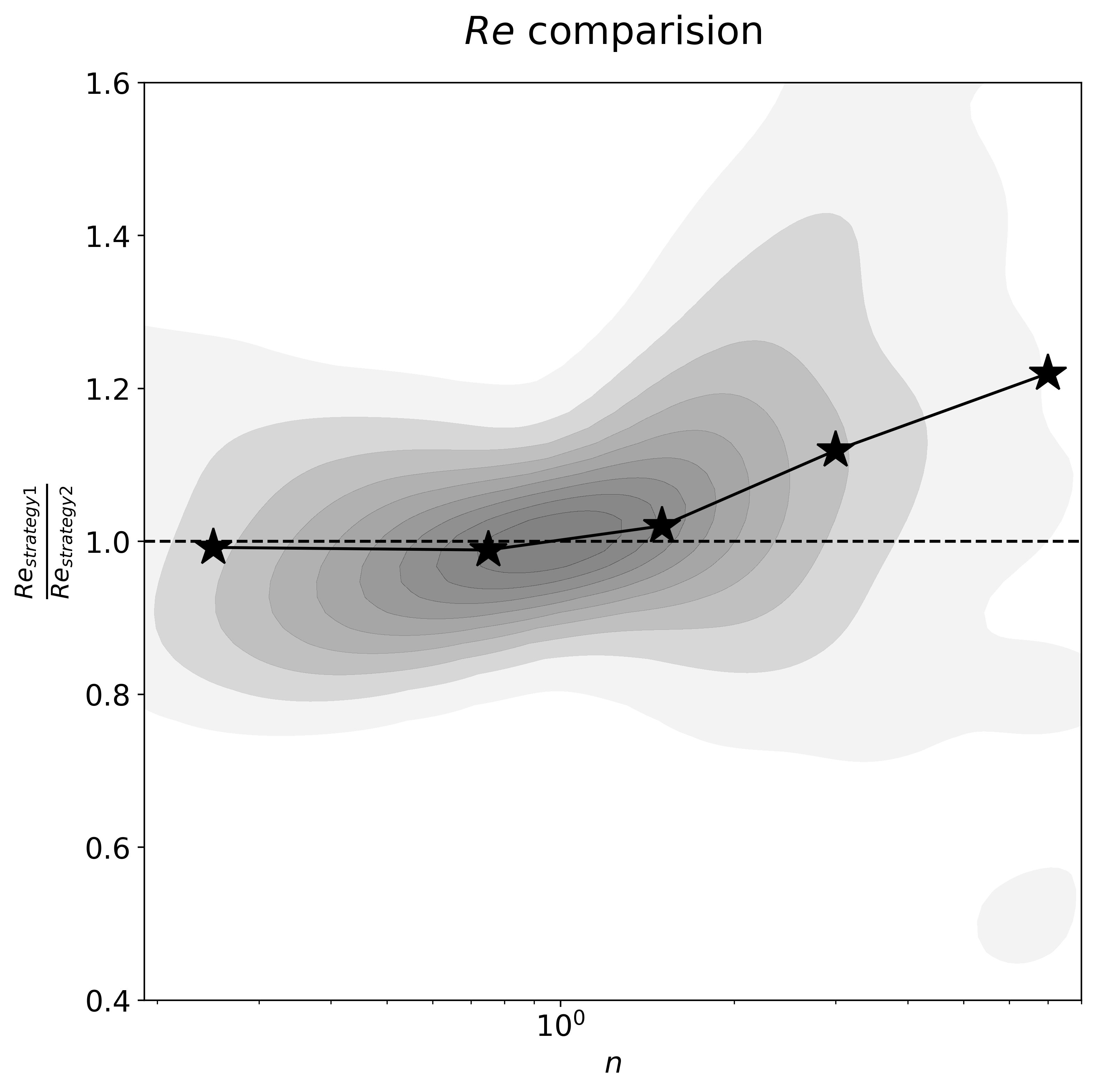}}
   \end{minipage}
\caption{Quality of our fits ($[\text{output}-\text{input}]/\text{input}$) of the S{\'e}rsic index and effective radius vs. input S/N, obtained by fitting 10000 mock galaxies, for strategy 1 (upper left and upper right panels, fitting all parameters) and strategy 2 (lower left panel, fixing S{\'e}rsic index to one). 
Mock galaxies are split into four different input effective radius bins. Data points are color-coded by their input S{\'e}rsic index. 
Sliding medians and the corresponding uncertainties (16th and 84th percentile) along the S/N are marked by black circles. 
25\% uncertainties are labeled by horizontal gray lines. 
Vertical color lines mark S/N of 40 and 10 for strategy 1 and strategy 2, respectively. 
For each input effective radius bin, the line charts above show the failing rate of GALFIT. In the lower right panel, we compare the ratio between the effective radius of strategy 1 and strategy 2 as a function of the input S{\'e}rsic index, with contours exhibiting the distribution and stars showing the sliding median.
}
\label{galfit-simulation}
\end{figure*}

In Fig. \ref{galfit-simulation}, we present the results of our simulation. The metric used to evaluate the goodness of fitting is the relative error of the output measurement (i.e., $(\text{output}-\text{input})/\text{input}$), and we studied the evolution of this quantity as a function of the S/N. Additionally, we gauged the goodness of these fits by studying the failure rate of GALFIT (fraction of the fits flagged as bad or non-fitting) as a function of S/N. To enhance clarity, we binned the data by their input effective radius and color-coded them based on their input S{\'e}rsic index. Because we are using optical structural parameters to generate a realistic set of mock galaxies, the number in each effective radius bin is not constant. For example, sources with a small effective radius (i.e., $<3$ pixels) are relatively rare.

By comparing the top two panels of Fig. \ref{galfit-simulation}, it is evident that estimating the S{\'e}rsic index of a galaxy is more challenging than determining its effective radius, likely due to the fact that a small change in the effective radius can be compensated for by a substantial change in the S{\'e}rsic index while maintaining a similar $\chi^2$. 
For strategy 1, we observe that down to S/N $\sim40$, the S{\'e}rsic index is well constrained by GALFIT, with the uncertainty staying below 50\%, even for very small sources. However, for galaxies with S/N below 40, the inferred S{\'e}rsic index appears quite dispersed around its true value and even systematically biased toward higher values, especially for sources with a small effective radius. Meanwhile, sources with S/N $<40$ show a high probability of failure, at a rate of even more than 50\% for very small sources. All of these suggest that the S{\'e}rsic index derived from strategy 1 is not reliable for galaxies with S/N $<40$. Consequently, we decided to adopt the results of the GALFIT fit of the S{\'e}rsic index (strategy 1) only for galaxies with S/N $\geq40$. 

To extend our structural analysis to galaxies with S/N lower than 40, we turned ourselves to strategy 2, in which the S{\'e}rsic index is fixed to the median value of the group. Fixing S{\'e}rsic index to 1, or 0.7, in strategy 2 is reasonable, considering that most SFGs exhibit disk-like structures with an approximately exponential light profile \citep[see Sect. \ref{Sersic index}; also][]{vdw2014, Paulino-Afonso2017}.
Compared with results from strategy 1 for the effective radius, we find that the GALFIT fits turn out to be better constrained (i.e., median and uncertainties of $(\text{output}-\text{input})/\text{input}$ are within 20\%) all the way down to S/N $\sim10$. Besides, by fixing the S{\'e}rsic index, GALFIT's failure rate significantly decreases by more than 30\% (from 65\% in strategy 1 at S/N $=40$ to 35\% in strategy 2), due to fewer degrees of freedom in the fit. 

Finally, at the bottom right of Fig. \ref{galfit-simulation}, we compare the effective radius inferred from strategy 1 and strategy 2, and hence restrict ourselves to sources with S/N $\geq40$. We notice that the fitting results of these two strategies are consistent with each other for galaxies with a S{\'e}rsic index that is less than 2. The observed underestimation in strategy 2 for sources with a relatively high S{\'e}rsic index is also seen as a systematic offset for galaxies in the bottom left panel of Fig. \ref{galfit-simulation}, especially for large galaxies. This artifact naturally comes from our assumption in Strategy 2 of fixing the S{\'e}rsic index to the value of one, while the intrinsic S{\'e}rsic index of these galaxies is significantly larger than one and in particular larger than two. 
Fortunately, the fraction of SFGs with a MIR S{\'e}rsic index greater than two is found to be marginal (10\%) in our S/N $\geq40$ sample (see Sect. \ref{Sersic index}); hence, the slight bias of strategy 2 that we see here should not have a significant impact on our main results.

In summary, for galaxies with S/N $\geq40$, we measured both their MIR effective radius and S{\'e}rsic index using strategy 1. With strategy 2, we extended our structural analysis to S/N $>10$ by fixing galaxies' S{\'e}rsic index to the median value observed in the S/N $\geq40$ sample from strategy 1 (i.e., $n_\text{MIR}\sim0.7$; see Sect. \ref{Sersic index}). We note that for parameters other than S{\'e}rsic index and effective radius, we find no significant difference in terms of fitting results between strategy 1 and strategy 2. It should be noted that galaxies in strategy 2 are always included in strategy 1. In the following sections, we present and discuss only the structural parameters deduced from one of these strategies, without mixing them, so that the analysis remains self-consistent.

Finally, we note that the errors of the structural parameters directly output by GALFIT are very small, with a median of only a few percent for the S{\'e}rsic index and effective radius, respectively. 
Such small errors are about three times smaller than the ones we deduced from our simulations by measuring the dispersion of the (output-input)/input distribution. Consequently, in what follows, we have used the errors produced by GALFIT, multiplied by a factor of three, as the uncertainties on the measured structural parameters of our galaxies.

\subsection{Main sequence of the MIR sample}
\label{Final sample}
\begin{figure*}
   \centering
   \includegraphics[width=16cm, height=13cm]{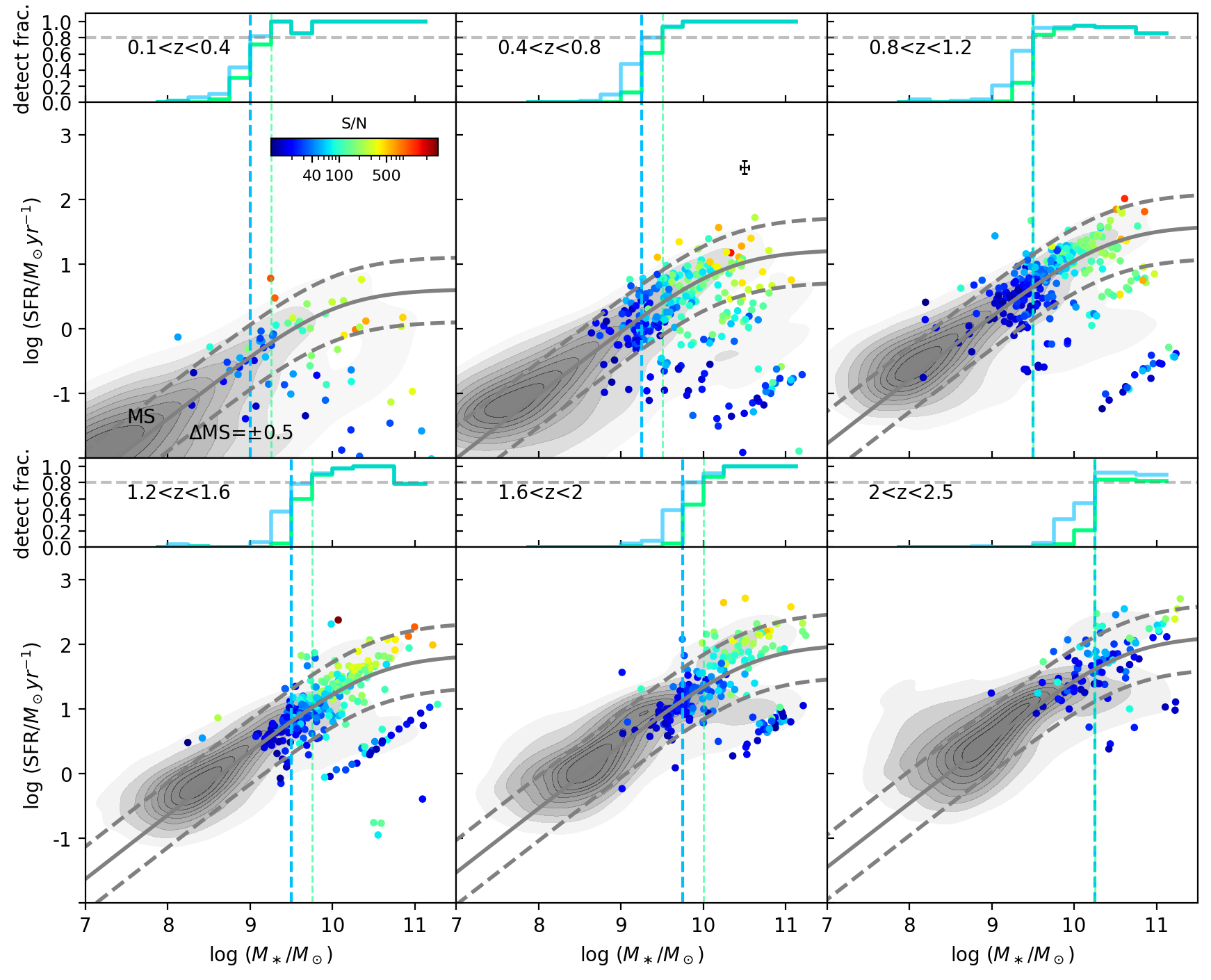}
    \caption{SFR vs. stellar mass for the CANDELS$\times$MIRI strategy 2 sample, grouped into six redshift bins from 0.1 to 2.5. The color of each point represents the S/N in MIR. Gray contours indicate the distribution of CANDELS galaxies falling on the PRIMER MIRI field of view. The solid gray curves are the main sequence of SFGs from \citet{Popesso2023}. The SFR is normalized according to the median redshift in each bin by maintaining the distance to the main sequence. The dashed gray lines are $\pm0.5$ dex from the main sequence. The green and blue histograms above exhibit the detection fraction of SFGs with $\Delta MS>-0.5$ for strategy 1 (S/N $\geq40$) and strategy 2 (S/N $\geq10$)  samples, respectively. The dashed vertical blue and green lines above display the 80\% stellar mass completeness limits of these samples (compared with CANDELS). The median error bar is shown in black.}
    \label{mass-SFR-total}
\end{figure*}

In Fig. \ref{mass-SFR-total}, we show the SFR as a function of the stellar mass for the whole sample of SFGs detected by MIRI with counterparts in the rest-frame optical, in six redshift bins between 0.1 and 2.5. The SFRs shown here are renormalized according to the localization of the main sequence at the median redshift of each bin; that is, the renormalized SFR is not the true SFR of the individual galaxy, but reflects the galaxy's distance from the main sequence at its redshift, so that the overall distribution of galaxies' $\Delta MS$ ($\equiv\text{log}\text{ [SFR]}-\text{log}\text{ [SFR}_{MS}$]) is maintained. We adopted the parametrization of the galaxy main sequence from \cite{Popesso2023}, who compiled a collection of literature measurements in the redshift range between 0 and 6.
The alignment of MIRI detected galaxies within $\pm0.5$ dex from the main sequence suggests that most of them are typical SFGs. Our sample also contains 36 ($\sim5\%$) galaxies (strategy 2) that exhibit extreme star formation activities and that are thus positioned well above the main sequence ($\Delta MS>0.5$; so called starburst galaxies). This fraction of starburst galaxies is consistent with the literature \citep{Rodighiero2011, Schreiber2015}. To further constrain the sample into pure SFGs, in the rest of the analysis we restricted the sample to galaxies with $\Delta MS \geq-0.5$ (effectively excluding few MIRI-detected QGs, $<10\%$). We note that the ``artificial'' stripe distribution of QGs is due to the SFH grid configuration that we used in the SED fit. This limited grid has no impact on our subsequent analyses of SFGs.
On the top panel of each main sequence plot, we calculated the completeness of our sample of pure SFGs with robust MIR morphological measurements by comparing their number with the total number of $\Delta MS \geq -0.5$ SFGs in the CANDELS catalog that are covered by our MIRI maps.
We find that, using CANDLES as a benchmark, our sample is 80$\%$ complete down to a stellar mass of $10^{9}$, $10^{9.25}$, $10^{9.5}$, $10^{9.5}$, $10^{9.75}$, and $10^{10.25}$ $\textit{M}_{\odot}$ at $z\sim0.25, 0.6, 1.0, 1.4, 1.8,$ and $2.2$ for strategy 2. These stellar mass completeness limits increase to $10^{9.25}$, $10^{9.5}$, $10^{9.5}$, $10^{9.75}$, $10^{10}$, and $10^{10.25}$ $\textit{M}_{\odot}$ for the same redshift bins when we are using strategy 1.
In the rest of our analysis, we restricted our sample to galaxies above these stellar mass limits and deemed it our final sample. 

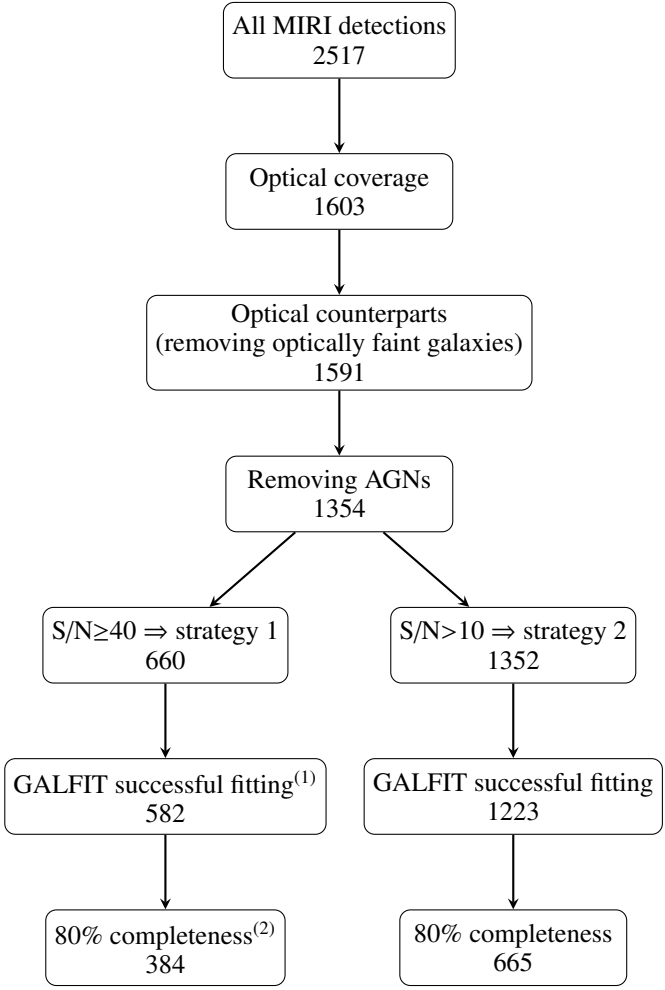
\begin{figure}
\centering
\tikzstyle{startstop} = [rectangle, rounded corners, 
minimum width=3cm, 
minimum height=1cm,
text centered, 
draw=black, 
fill=white!30]
\tikzstyle{arrow} = [thick,->,>=stealth]

\begin{tikzpicture}[node distance=2cm]

\node (input1) [startstop, align=center] {All MIRI detections 
\\
2517};
\node (input12) [startstop, below of=input1, align=center] {Optical coverage
\\
1603};
\node (input2) [startstop, below of=input12, align=center] {Optical counterparts
\\ (removing optically faint galaxies)
\\
1591};
\node (input3) [startstop, below of=input2, align=center] {Removing AGNs
\\
1354};
\node (input4) [startstop, below of=input3, xshift=-2.3cm, align=center] {S/N$\geq$40 $\Rightarrow$ strategy 1
\\
660};
\node (input5) [startstop, below of=input3, xshift=+2.3cm, align=center] {S/N$>$10 $\Rightarrow$ strategy 2
\\
1352};
\node (input41) [startstop, below of=input4, align=center] {GALFIT successful fitting$^{\text{(1)}}$
\\
$582$};
\node (input51) [startstop, below of=input5, align=center] {GALFIT successful fitting
\\
$1223$};
\node (input42) [startstop, below of=input41, align=center] {80\% completeness$^{\text{(2)}}$
\\
$384$};
\node (input52) [startstop, below of=input51, align=center] {80\% completeness
\\
$665$};
\draw [arrow] (input1) -- (input12);
\draw [arrow] (input12) -- (input2);
\draw [arrow] (input2) -- (input3);
\draw [arrow] (input3) -- (input4);
\draw [arrow] (input3) -- (input5);
\draw [arrow] (input4) -- (input41);
\draw [arrow] (input5) -- (input51);
\draw [arrow] (input41) -- (input42);
\draw [arrow] (input51) -- (input52);

\end{tikzpicture}
\caption{Summary of our sample selection steps. (1) Flags from GALFIT have a value of good or acceptable; (2) Galaxies above our 80\% mass completeness limits as defined in Sect. \ref{Final sample}. All galaxies in strategy 1 are included in strategy 2.}
\label{flow-chart}
\end{figure}

Our final strategy 1 sample provides the full structural parameters (S{\'e}rsic index, effective radius, position angle, axis ratio, position, and magnitude) of 384 galaxies, representative of the massive population of SFGs at $z<2.5$. Our final strategy 2 sample provides, by fixing galaxies’ S{\'e}rsic index to 0.7, the partial structural parameters (effective radius, position angle, axis ratio, position, and magnitude) of 665 galaxies, representative of the intermediate and massive stellar mass population of SFGs at $z<2.5$. Figure \ref{flow-chart} summarizes the selection procedures of these two final samples and the number of galaxies involved in each step.

\section{Results}
\label{results}
In this section, we investigate the differences between the morphologies of galaxies observed in the rest-frame optical after color gradient correction and those seen in the rest-frame MIR. 
Our goal is to investigate whether there are potential correlations between the structural parameters of galaxies and their physical properties, and in particular whether there is, as in the rest-frame optical, a clear mass--size relation in the rest-frame MIR.
By pursuing this study, we seek to decipher how stellar and star-forming components are distributed inside galaxies and how they are connected with each other. This enables us to better understand the mechanisms that govern the structural evolution of SFGs, as well as the halt in their star formation.

For ease of reading, in the following sections we refer to our rest-frame optical after color gradient correction and rest-frame MIR measurements (e.g., effective radius, S{\'e}rsic index, etc.) as simply optical and MIR measurements. Moreover, when we refer to the size of a galaxy, we always mean its effective radius.
\subsection{Axis ratio and position angle}
\begin{figure}[ht]
    \begin{minipage}[c]{0.95\columnwidth}
        \centering
        \center{\includegraphics[width=0.99\linewidth]{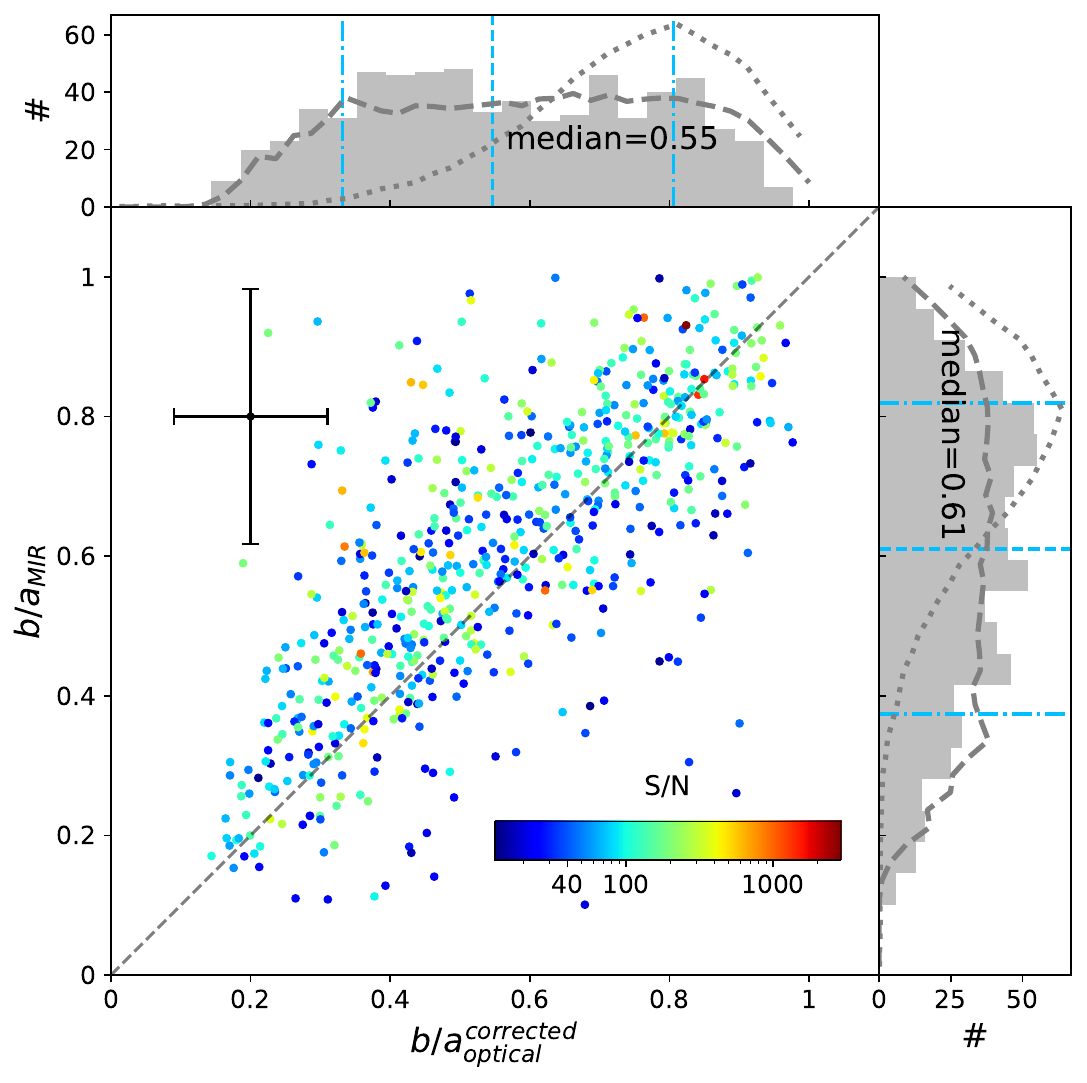}}
   \end{minipage}
   \begin{minipage}[c]{0.95\columnwidth}
        \centering
        \center{\includegraphics[width=0.99\linewidth]{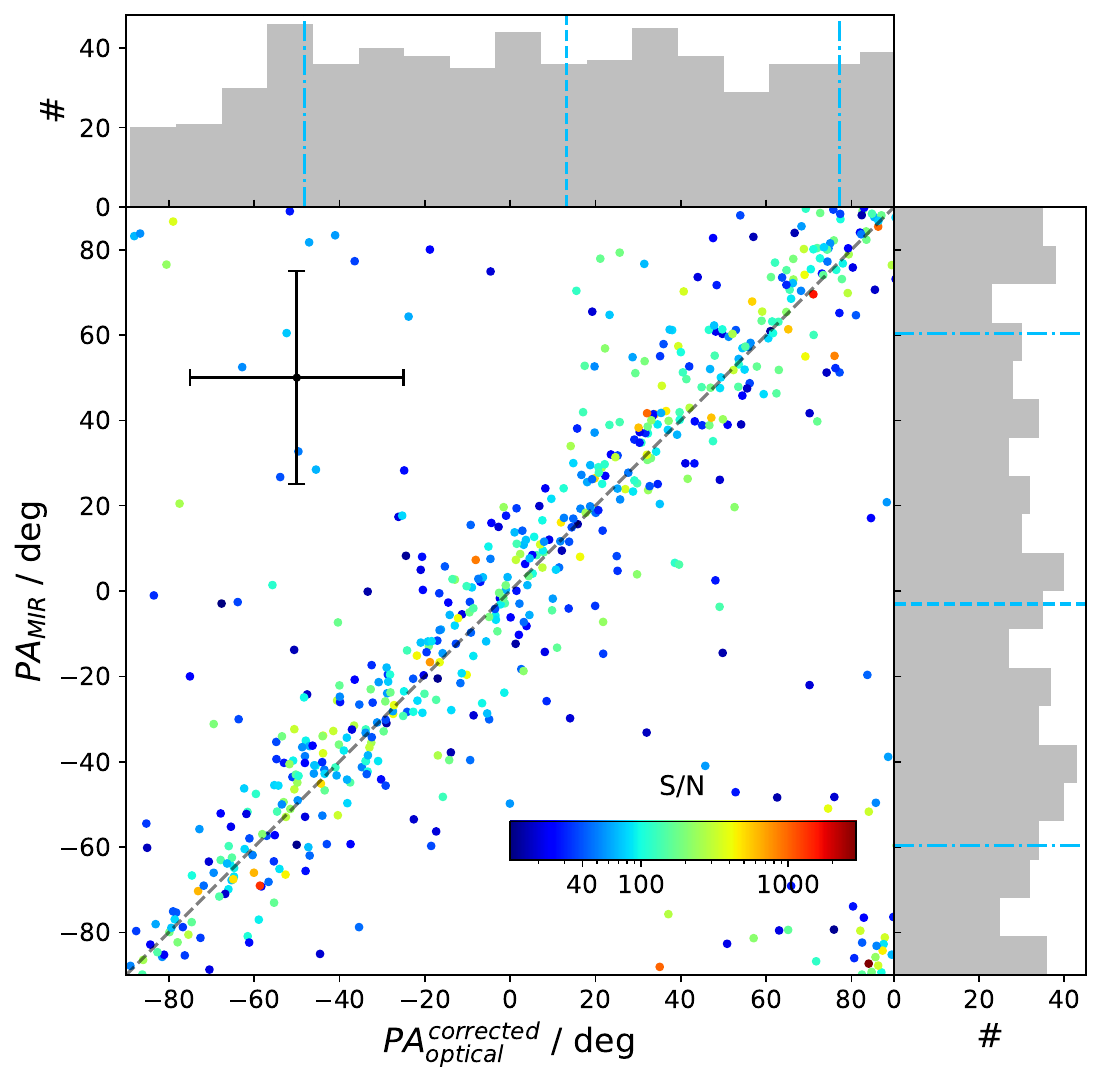}}
   \end{minipage}
   \caption{Axis ratio and position angle comparison between the optical and MIR measurements for our SFGs using strategy 2. Galaxies are color-coded by their S/N. The dashed gray line shows the 1-to-1 line. Sidebars exhibit the corresponding distributions. The median and 16th/84th percentiles of these distributions are illustrated by the dashed and dash-dotted blue lines, respectively. The median error bar is shown in black. Dashed and dotted gray curves are axis ratio distributions of local spiral and elliptical galaxies, respectively, referred from \citet{Rodriguez2013}. For the position angle, galaxies that are clustered in the lower right-hand corners of the graph are not really outliers. This appears due to the fact that at these high position angles, a galaxy can be fit indifferently with a 90 degree or -90 degree position angle.}
   \label{optical-MIRI-loc}
\end{figure}
In Fig. \ref{optical-MIRI-loc}, we compare the axis ratio and position angle observed in the  optical and MIR, both inferred using our strategy 2 (i.e., for galaxies with S/N $>10$). Both the optical axis ratios and MIR axis ratios have a median value of around 0.5 (the MIR axis ratio is on average slightly larger than optical one, suggesting a more bulge-like star-forming core in these galaxies), and show a relatively flat distribution. In the local Universe, elliptical QGs typically exhibit a skewed Gaussian distribution of axis ratios with a peak around 0.7 or higher; in contrast, spiral SFGs often display a more uniform or flat distribution, centered around 0.5 \citep{Rodriguez2013}. Consequently, the flat axis ratio distributions observed here, together with the fact that the S{\'e}rsic index distributions in the optical and MIR peak around $1$ and $0.7$, respectively (see Sect. \ref{Sersic index}), indicate that both the stellar and star-forming components of most SFGs have disk-like structures. We note that previous works have found that while most massive SFGs at low redshift ($z<1.5$ and $M_\ast>10^{10}M_\odot$) have disk-like structures, at higher redshift or low mass the SFGs have more irregular structures \citep[e.g.,][]{Zhang2019}. However, we do not observe in our data any clear evolution of the axis ratio distribution with reshift and/or stellar mass, probably due to limited statistics. We defer a more detailed exploration of this trend to future work.

At the same time, we find that there is a very good correlation of the axis ratio and the position angle measured between the optical and MIR, closely following the 1-to-1 line. We also find the median astrometric offset between the  optical and MIR centroids is just $0\farcs07$ (i.e., 500 pc at $z\sim1$, compared with $0\farcs6$ angular resolution of the MIRI \textit{F1800W} map). All these imply that in our SFGs, not only are the stellar and star-forming components disk-like, but they also align well with each other. We note that from a technical point of view, the good agreement between the optical and MIR position angles instills good confidence in our MIR profile fits in general, as it implies that the shape (minor and major axis), and therefore the size of the MIR component, has been successfully measured. Restricting this analysis to our strategy 1 sample would not change any of these conclusions, but would simply lower the statistic.

\subsection{S{\'e}rsic index}
\label{Sersic index}
\begin{figure}
\centering
\begin{minipage}[c]{0.95\columnwidth}
\center{\includegraphics[width=0.99\linewidth]{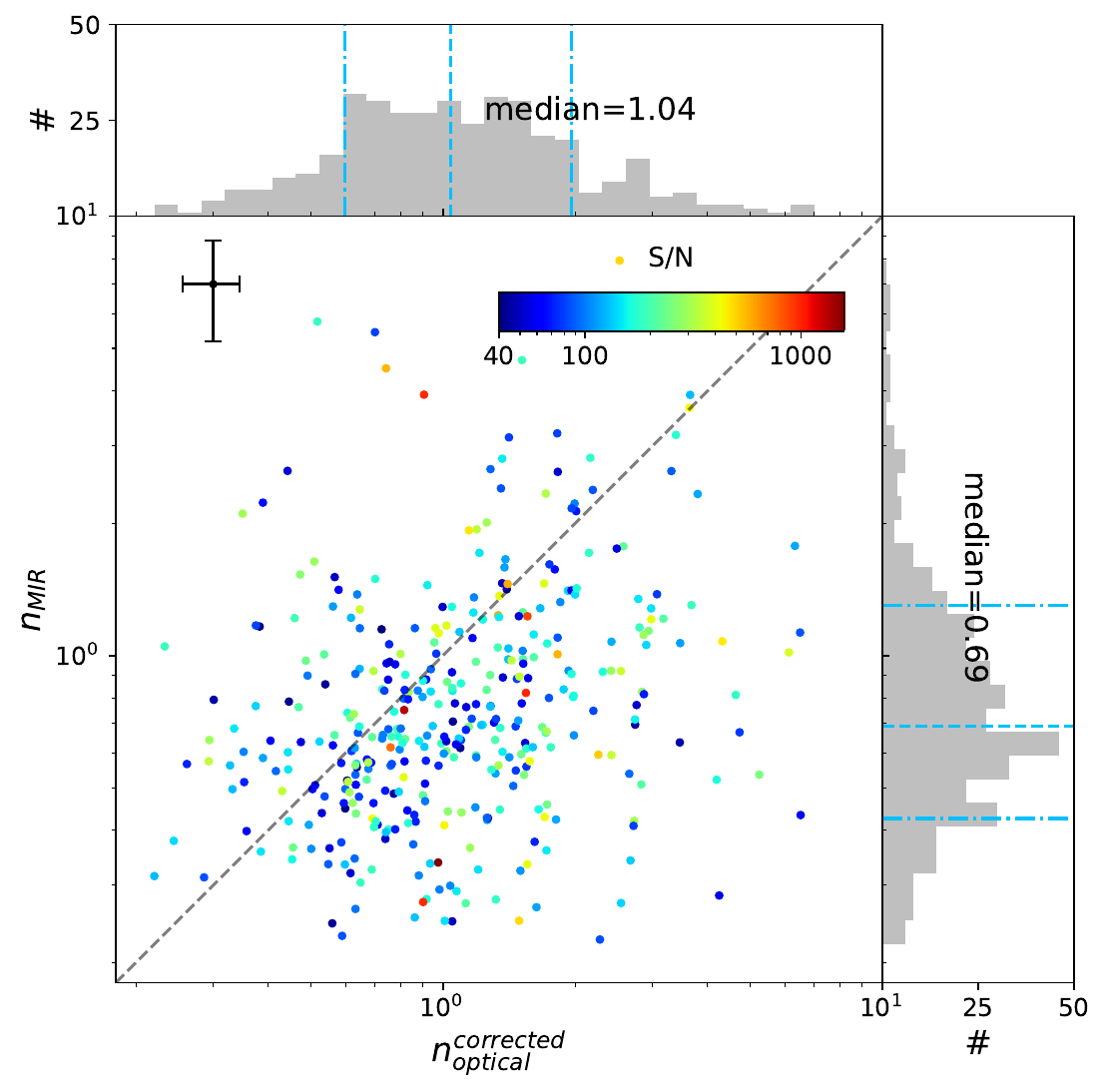}}
    \caption{Optical S{\'e}rsic index versus MIR S{\'e}rsic index inferred from strategy 1, color-coded by their S/N. The median error bar is shown in black in the upper left. The dashed gray line shows the 1-to-1 line. The corresponding distributions are shown in the sidebars, with the median value and 16th/84th percentiles indicated with dashed and dash-dotted blue lines.}
    \label{optical-MIRI-n}
\end{minipage}
\end{figure}
Through strategy 1, we obtained robust measurements of the S{\'e}rsic index for 384 galaxies, with $z<2.5$ and $\Delta MS>-0.5$. In Fig. \ref{optical-MIRI-n}, we compare the distribution of their MIR S{\'e}rsic index with their optical S{\'e}rsic index. There is a weak correlation between them, with a Pearson correlation value of 0.27. The median S{\'e}rsic index value for MIR and optical are $0.69^{+0.66}_{-0.26}$ and $1.04^{+0.93}_{-0.50}$, respectively (the range corresponds to the 16th and 84th percentiles of the distribution). The typical error on the optical and MIR S{\'e}rsic index is around 20\%, which is smaller than the dispersion around the 1-to-1 line observed in Fig. \ref{optical-MIRI-n}, as well as smaller than the 16th to 84th percentile ranges given above. A median S{\'e}rsic index value of one in the optical implies that the stellar components of these galaxies have an exponential light profile, which, combined with their flat axis ratio distribution, demonstrates that these stellar components resemble local disks.
In slight contrast, the MIR emission of these galaxies follows a disk-like structure according to their axis ratio distribution but with a flatter light profile, between that of a Gaussian and exponential profile, as is also found in \citet{Shen2023} and \citet{Magnelli2023}. In order to investigate whether these distributions of S{\'e}rsic index in the MIR and optical could, however, have been drawn from the same parent distribution, we performed a Kolmogorov-Smirnov test. We find a statistic-value of 0.25 and a $p$-value $\ll0.05$, both implying that the MIR and optical S{\'e}rsic index distributions are different from each other. Although subtle, these differences are therefore statistically robust; that is, the dust-obscured star-forming components of SFGs have a flatter light profile than the corresponding stellar components. This difference is probably not due to the different angular resolution between the optical and MIRI \textit{F1800W} images used here, since \citet{Shen2023} also found $n_{MIR}\sim0.7$ for their sample, for which they used data with two times better angular resolution offered by the MIRI \textit{F1000W} image. This flatter MIR light profile could be indicative of a clumpier or patchier star-forming disk, although ALMA observations down to a resolution of hundreds of parsecs found no evidence for significant dust-obscured star-forming clumps \citep{Cibinel2017, Ivison2020}. Further observations with higher sensitivity and resolution are required to confirm this.
As a final comment on this distribution of S{\'e}rsic index, we note that 92\% of the SFGs in our strategy 1 sample have a S{\'e}rsic index below two in the MIR. This implies that the structural parameters deduced in our strategy 2 sample by fixing the S{\'e}rsic index to 0.7 are robust within $10\%$ (see Fig. \ref{galfit-simulation}). 

\begin{figure*}
    \centering
    \includegraphics[width=16cm]{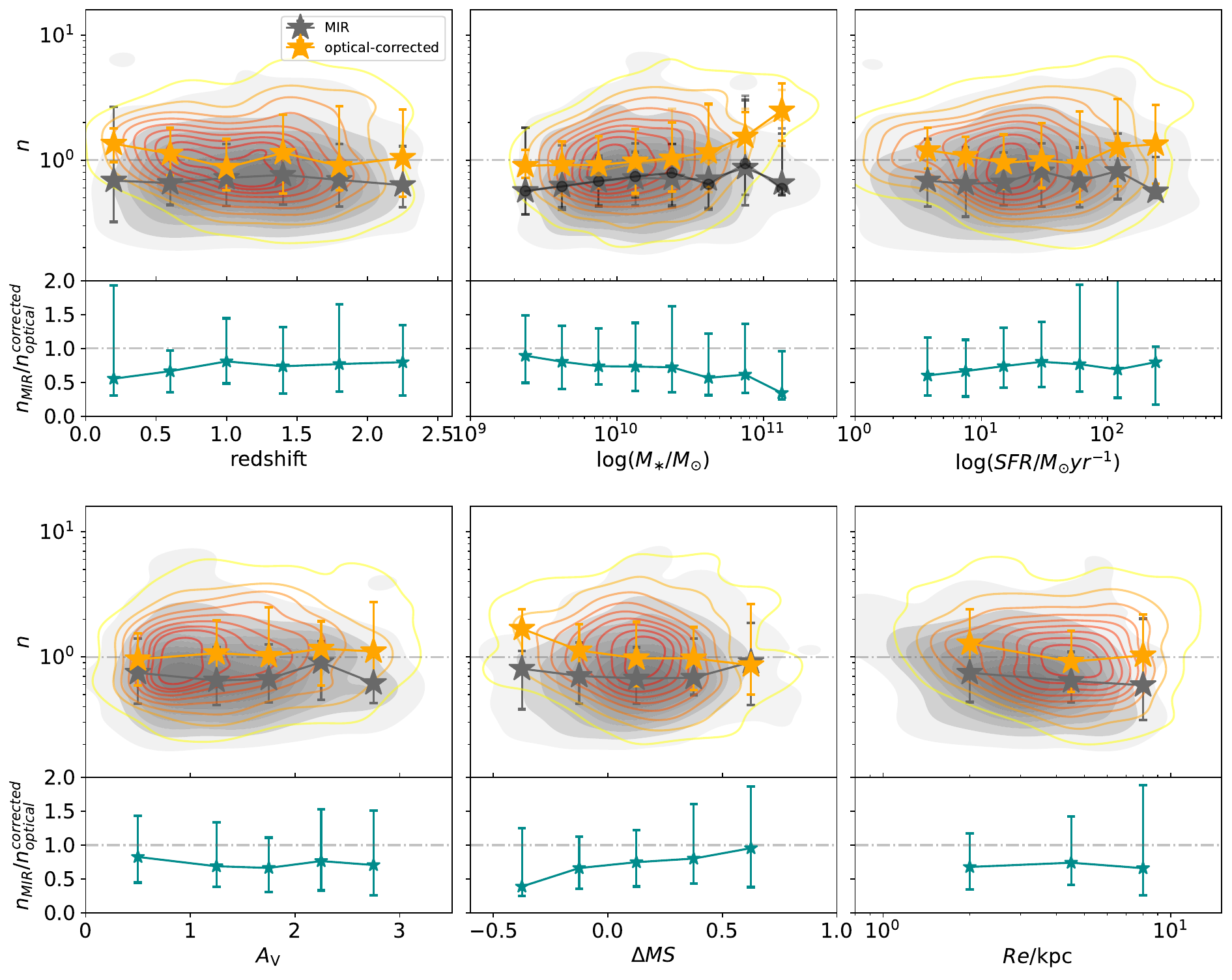}
    \caption{S{\'e}rsic index (from strategy 1) as a function of various galaxy properties (redshift, stellar mass, SFR, dust attenuation, distance to main sequence, $\Delta MS$, and MIR effective radius). Gray and yellow contours are the underlying distribution for the MIR and optical measurements, respectively. Gray and yellow stars are the sliding median value, with 16th and 84th percentile as error bars. The median error bars of different physical properties of galaxy are shown in black. The bottom frame exhibits the S{\'e}rsic index ratio between MIR and optical measurements as a function of the same physical properties. In the stellar mass panel, gray and yellow circles are selected galaxies with $\Delta MS>0$.}
    \label{n_change}
\end{figure*}

In order to investigate the correlation between the S{\'e}rsic index and other physical properties of our SFGs, we compare in Fig. \ref{n_change} the MIR S{\'e}rsic index, the optical S{\'e}rsic index, and their ratio ($n_\text{MIRI}/n_\text{optical}$) with various galaxy properties: redshift, stellar mass, SFR, dust attenuation index, $A_\text{V}$, $\Delta MS$, and effective radius. We find that there is no evidence for both the optical and MIR S{\'e}rsic index evolving with redshift.
There is also no significant correlation between the S{\'e}rsic index and $A_\text{V}$, in both the MIR and optical. 
With the effective radius, we find a slight anticorrelation in both the optical and MIR. Because this trend occurs in images with vastly different angular resolution ($0\farcs1$ versus $0\farcs6$), it suggests that it is not due to a degeneracy inherent to GALFIT, but rather implies that the stellar and star-forming components of the smallest galaxies are more centrally concentrated.

We observe that the S{\'e}rsic index decreases slightly with $\Delta MS$ in the optical. This indicates that when galaxies start to quench, and thus move below the main sequence, their stellar light resembles more and more that of QGs with a high S{\'e}rsic index. We do not observe the same anticorrelation between $\Delta MS$ and the MIR S{\'e}rsic index. This suggests that for these SFGs, even on the way to quiescence, their star-forming components still maintain the shape of an exponential disk. 

Finally, we detect a positive correlation between S{\'e}rsic index and stellar mass in the optical. We do not find the same strong trend in the MIR. To ensure that part of this correlation could not be affected by the negative correlation between the S{\'e}rsic index and $\Delta MS$ mentioned earlier, we restricted our sample to those with $\Delta MS>0$.
After doing this, we find that the correlation between S{\'e}rsic index and stellar mass in the optical still exist. This indicates that when galaxies become more massive, their stellar components become more centrally concentrated, while their dusty star-forming components still maintain a relatively flat light distribution. This positive stellar mass trend in optical, together with the positive relation between stellar mass and the SFR (main sequence), results in more SFGs (higher SFR) tending to have a higher optical S{\'e}rsic index, as is seen in Fig. \ref{n_change}.

All the above can be summarized and explained as follows: on the one hand, the optical features of our galaxies are consistent with previous studies; that is, the stellar components of SFGs consist of two separate parts, one disk ($n\sim1$) and one bulge ($n\sim4$), and the bulge-to-total mass ratio (total = disk + bulge) increases strongly with stellar mass \citep{Khochfar2011, Bluck2014}. Therefore, as the stellar mass increases, the optical S{\'e}rsic index rises from $n\sim1$ (B/T~$\sim0$) to $n\sim4$ (B/T~$\sim1$). On the other hand, the star-forming components remain mostly in a disk-like structure at all stellar masses, suggesting that the bulge might not form in situ in a secular slow process but rather probably in violent out-of-equilibrium phases. This point is discussed in more detail in Sect. \ref{discussion}. To complement this scenario, we now turn our attention to the size evolution of the stellar and star-forming components of these galaxies.

\subsection{Effective radius}
\label{Effective radius}
\begin{figure}
\begin{minipage}[c]{0.95\columnwidth}
\center{\includegraphics[width=0.99\linewidth]{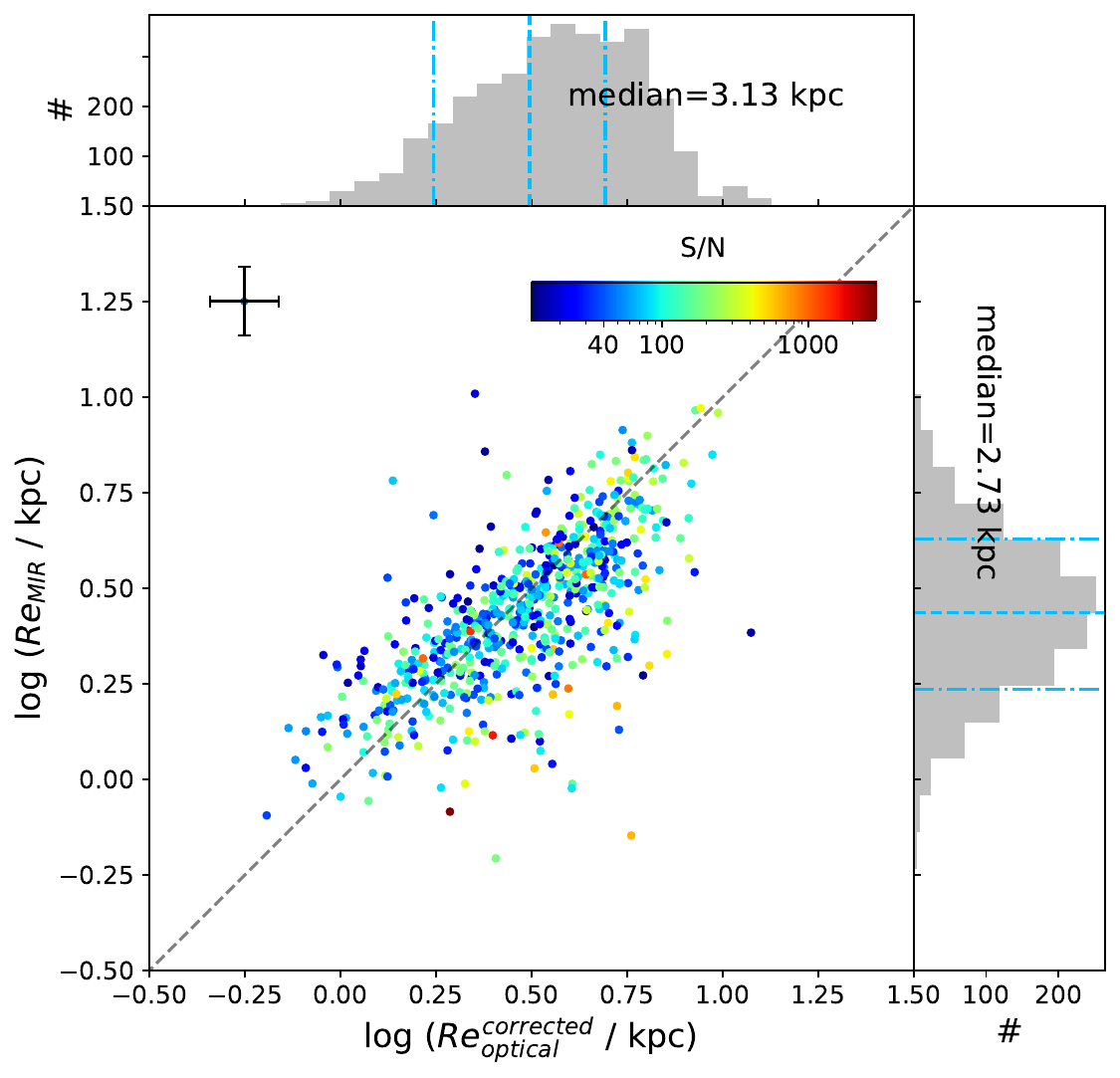}}
   \caption{Comparison of the optical and MIR effective radius of the SFGs in our strategy 2 sample. Galaxies are color-coded by their S/N. The dashed gray line is the 1-to-1 line. The median error bar is shown in black in the upper left. Sidebars exhibit the corresponding distributions, with the median value and 16th/84th percentiles showed as dashed and dash-dotted blue lines, respectively.}
   \label{optical-MIRI-re}
   \end{minipage}
\end{figure}
Through strategy 2, we obtained robust measurements of the MIR effective radius for 665 galaxies, which constitutes a mass-complete sample of SFGs down to $10^{9.75} \textit{M}_{\odot}$ at $z\sim2$.
In Fig.~\ref{optical-MIRI-re}, we compare the MIR effective radius with the optical ones, color-coded by their S/N.
There is a strong correlation between these optical and MIR measurements, with a Pearson correlation value of 0.67, suggesting strong consistencies between the size of the stellar and star-forming components in these galaxies. We find that the median value for the optical effective radius is $3.13^{+1.80}_{-1.37}$ kpc and the median value for MIR effective radius is $2.73^{+1.51}_{-1.02}$ kpc (the ranges correspond to the 16th and 84th percentiles of the distribution).

\begin{figure*}
    \centering
    \includegraphics[width=16cm]{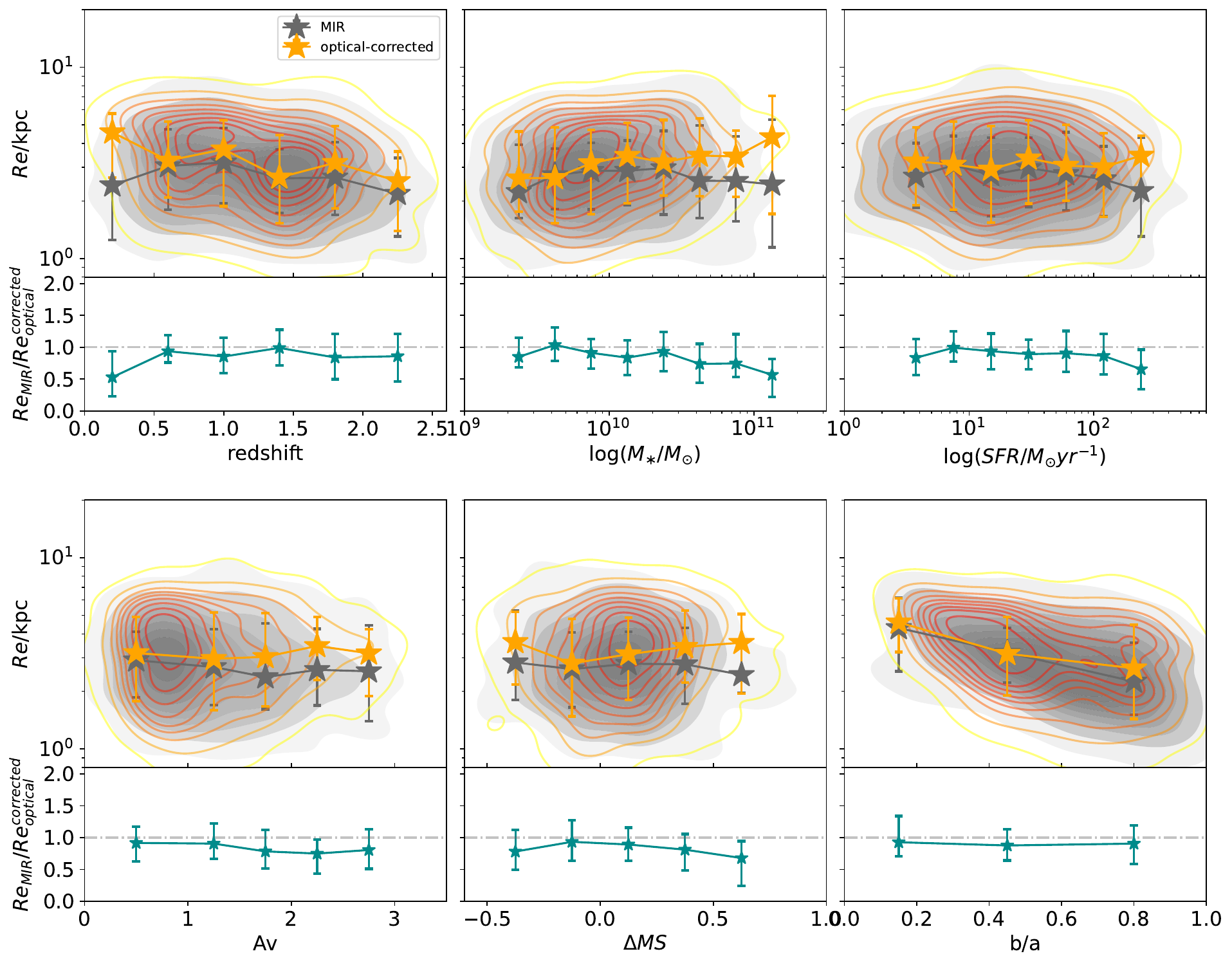}
    \caption{Effective radius inferred from our strategy 2 sample, vs. various galaxy properties (redshift, stellar mass, SFR, dust attenuation, distance to main sequence, $\Delta MS$, and MIR axis ratio). Gray and yellow contours mark the underlying distribution for MIR and optical measurements. Gray and yellow stars correspond to the median values, with 16th and 84th percentiles as error bars. The median error bars of different physical properties of galaxy are shown in black. In the bottom part of each panel, we display the effective radius ratio between MIR and optical measurements as a function of galaxy properties.}
    \label{re_change}
\end{figure*}

In Fig. \ref{re_change}, we compare the optical, MIR effective radius and their ratio (see more details in Fig.\ref{appendix12}) with galaxies' physical properties. 
Neither the MIR nor the optical effective radius correlate significantly with $\Delta MS$, SFR, and dust attenuation, $A_{\text{V}}$.
As for the axis ratio, we observe a pronounced anticorrelation whereby smaller galaxies tend to be more round. Given that the same anticorrelation is seen in both optical and MIR, this cannot be simply ascribed to some inherent degeneracy in GALFIT (HST-\textit{F160W} have six times better angular resolution than MIRI-\textit{F1800W}). This anticorrelation with the axis ratio implies that the stellar and star-forming components of the smallest galaxies (low-mass blue nuggets or compact SFGs; see Sect. \ref{three-classification}) are not only more centrally concentrated (see Sect. \ref{Sersic index}) but also more round, as would be expected if we were witnessing the growth of the stellar bulge in galaxies.
As for the redshift, we observe a negative correlation for both optical and MIR effective radius, except at very low redshift where our sample size is, however, not large enough to provide a robust statistic. This slight redshift evolution reflects that of the mass--size relation and is further investigated in Sect. \ref{Mass-size relation}. 

As for the stellar mass, we observe that the optical effective radius is quite consistent with the MIR effective radius up to $\textit{M}_{\ast} \sim 10^{10.5} \textit{M}_{\odot}$. In low- and intermediate-mass SFGs, the stellar and star-forming components thus have very similar structures in terms of the S{\'e}rsic index, axis ratio, and effective radius \citep[see also ][]{Magnelli2023}, and both stellar and star-forming components increase in size as the stellar mass increases. 
In contrast, at higher stellar masses ($\textit{M}_{\ast} > 10^{10.5} \textit{M}_{\odot}$), while the optical median effective radius of SFGs still increases slightly or reaches a plateau, the MIR effective radius of these massive SFGs is highly scattered, with some galaxies being extended and others being compact, but with an overall decrease in the median MIR sizes at these stellar masses. This suggests the emergence at these stellar masses of a large number of galaxies with compact star-forming cores embedded in larger stellar components.

Combining all the information obtained so far, the structural evolution that galaxies undergo as their mass increases becomes clearer: as stellar mass increases, the stellar components of SFGs increase in size and become increasingly bulge-dominated, while their star-forming components also increase in size, but mainly retain their disk-like structure. At high mass, there appears to be the emergence of compact star-forming components in galaxies whose stellar component is otherwise extended. Interestingly, we also find that the stellar and star-forming components of smaller galaxies are more ``bulgy'' (larger S{\'e}rsic index and axis ratio). To investigate in more detail the structural evolution of galaxies as they increase in mass, in the next section we examine the MIR and optical mass--size relations as a function of redshift.

\subsection{Mass--size relation}
\label{Mass-size relation}
\begin{figure*}
    \centering
    \includegraphics[width=17cm]{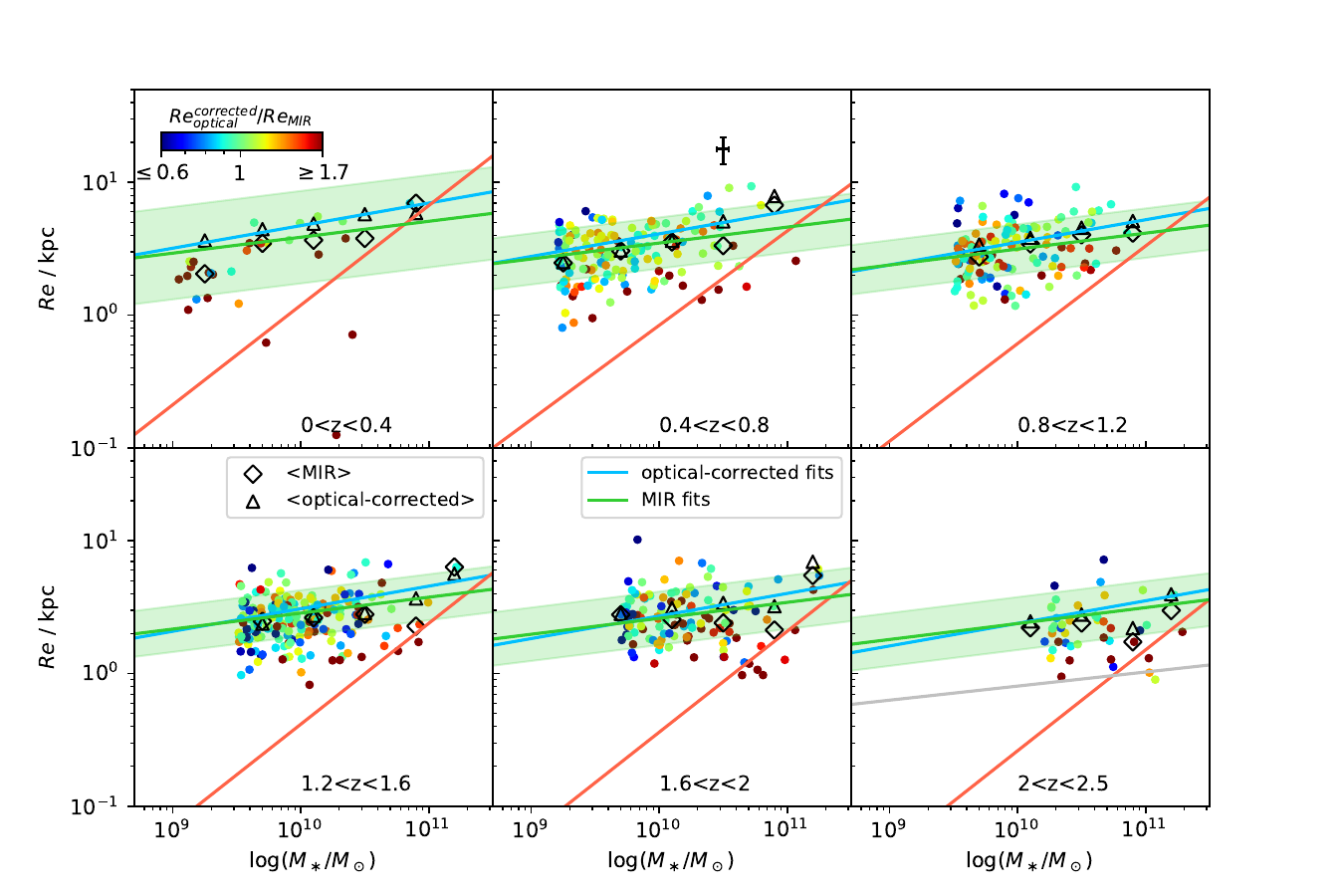}
    \caption{Effective radius vs. stellar mass for galaxies in our strategy 2 sample divided into six redshift bins. Circles are MIR size color-coded by their optical-to-MIR size ratio. Black diamonds and triangles represent the sliding median for MIR and optical measurements, respectively. The median error bar is shown in black in the upper middle panel. The solid green lines are our best-fitting mass--size relation for the MIR; the shaded regions are the corresponding $1\sigma$ uncertainties. The solid blue lines are our best-fitting optical mass--size relation. The solid red lines are the optical mass--size relation inferred in \citet{vdw2014} for QGs. The solid gray line is the mass--size relation inferred from \citet{CG2022} with ALMA for massive, high-redshift SFGs ($\textit{M}_{\ast}>10^{10.5}\textit{M}_{\odot}$, $z>2.5$). }
    \label{mass_size_relation}
\end{figure*}

\begin{figure*}
    \centering
    \includegraphics[width=15.5cm]{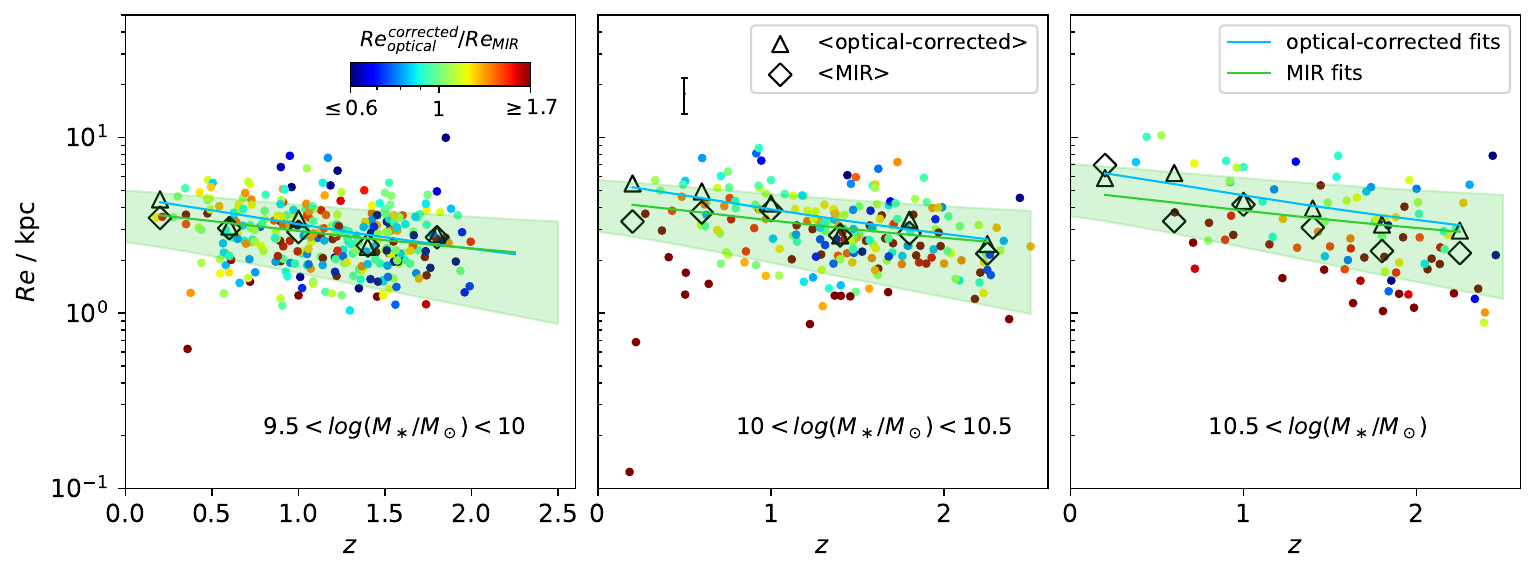}
    \caption{Redshift evolution of the MIR effective radius in three different stellar mass ranges. Circles are color-coded by their MIR-to-optical size ratio. The black diamonds are the sliding median for the MIR sizes. The median error bar is shown in black in the upper middle panel. The green lines show the best fitting results using Eq. \ref{eq4} and the shaded green regions are the corresponding scatter. Finally, the black triangles are the sliding median for the optical sizes of the same galaxies, while the blue lines show the best-fitting results for these sizes using Eq. \ref{eq4}. }
    \label{re_z}
\end{figure*}

As we saw in Sect. \ref{Effective radius}, the size of the stellar and star-forming components in SFGs appear to depend to the first order on their stellar mass, and to the second order on their redshift. To disentangle these two dependencies, we explore their MIR mass--size relation in distinct redshift bins in Fig. \ref{mass_size_relation}, and investigate their redshift evolution in Fig. \ref{re_z}. To obtain the best statistics and the greatest leverage in terms of stellar mass probed, we used our strategy 2 sample. 

We observe in Fig. \ref{mass_size_relation} a positive correlation between the stellar mass and the MIR effective radius of galaxies at all redshifts; that is, their star-forming component becomes more extended as their stellar mass increases. To better characterize this mass--size relation in the MIR, we followed the approach used in the optical \citep{Shen2003, vdw2014}, and adopted a single power-law function that assumes that galaxy effective radius adheres to a log-normal distribution:
\begin{equation}
\centering
\text{log }(Re\text{ / kpc})=\text{log }(A)+\alpha\text{ log }(\textit{M}_{\ast}\text{ / }5\times10^{10}\textit{M}_{\odot}),
\label{eq2}
\end{equation}
where $A$ is the intercept at the stellar mass of $5\times10^{10}\textit{M}_{\odot}$, and $\alpha$ is the slope of this power law relation. Unfortunately, since we are limited by the current number counts of MIRI detections, it is not possible to constrain the slope of the mass--size relation in each redshift interval independently, in particular for our lowest redshift bin (small comoving volume probed) and highest redshift bin (limited range of stellar mass probed). As an alternative, and motivated by the studies of the optical mass--size relation in \citet{vdw2014}, we assumed that the slope of the mass--size relation is independent of the redshift and that only its intercept evolves with the redshift. Furthermore, as is advocated in \citet{vdw2014}, we assumed that the redshift evolution of this intercept is better parameterized with the normalized Hubble parameter ($H_n\equiv H(z)/H(z=0)$) than with $(1+z)$. Indeed, it has been shown that the galaxy’s disk length is strongly correlated with the size of the dark matter halo that they are embedded in. In the Universe dominated by matter (i.e., $z\gtrsim1$), the size of the dark matter halo evolves tightly with the cosmological scale factor $(1+z)$ or $H_n$, as these two factors evolve very similarly in these epochs \citep{Dodelson2003}. However, in today’s dark-energy-dominated Universe, the size of the dark matter halo is found to be more correlated with the cosmological expansion rate \citep[i.e., $H_n$;][]{Dodelson2003}, which evolves much more slowly than $(1+z)$. Taking into account this redshift evolution of the intercept:
\begin{equation}
\centering
\text{log }(A)=\text{log }(B)+ \beta\text{ log }(H_n),
\label{eq3}
\end{equation}
and combining this with Eq. \ref{eq2}, we get:
\begin{equation}
\centering
\text{log }(Re\text{ / kpc})=\text{log }(B) + \alpha\text{ log }(\textit{M}_{\ast}\text{ / }5\times10^{10}\textit{M}_{\odot}) + \beta\text{ log }(H_n),
\label{eq4}
\end{equation}
where $B$ is now the intercept at the stellar mass of $5\times10^{10} \textit{M}_{\odot}$ when $z=0$, and $\beta$ is the slope of the redshift evolution.

The best fits of the MIR mass--size--redshift relation obtained using Eq. \ref{eq4} are shown in Figs. \ref{mass_size_relation} and \ref{re_z} and summarized in Table \ref{mass-size-fit-results}. We find that this simple log-norm function fits the median values of the MIR size of our MIRI galaxies relatively well. This is particularly true for low- and intermediate-mass galaxies. However, we observe some deviation from the best-fitting results for galaxies at high stellar masses. Although this deviation may be partly due to the small-number statistics in these parameter ranges, it suggests that the size distribution of the star-forming components of massive SFGs is shifted or skewed toward compact objects. \citet{vdw2014} also observe an asymmetric distribution in the case of the optical mass--size relation, but across all stellar mass and redshift ranges. They propose that these small galaxies might represent a transition phase between SFGs and QGs. 
\begin{table}
\caption{Fitting results.}
\label{mass-size-fit-results}
\centering
\renewcommand\arraystretch{1.5}
\begin{tabular}{c c c c}
\hline\hline
& log($B$) & $\alpha$ & $\beta$ \\  
\hline
   MIR & 0.69 (0.02) & 0.12 (0.01) & -0.44 (0.05) \\
   optical-corrected & 0.82 (0.02) & 0.17 (0.01) & -0.62 (0.04) \\
\hline
\end{tabular}
\tablefoot{Best-fitting results and the corresponding uncertainties (in parenthesis) of the mass--size--redshift relations using Eq. \ref{eq4}.}
\end{table}
\renewcommand\arraystretch{1}


In Fig. \ref{mass_size_relation}, we directly compare the mass--size relation in the MIR with the one in the optical, also fit using Eq. \ref{eq2} (see Table \ref{re_z}). Because both the MIR and optical relations (median values and fit using Eq. \ref{eq4}) were deduced from the same sample, this comparison is exempt from potential problems introduced by the combination of the small-number statistics and relatively large dispersion in the respective mass--size relations. We find that the size of the stellar and the star-forming components of low- to intermediate-mass galaxies are very similar. However, deviations of $10-30\%$ between the MIR and optical appear at the most massive end ($\textit{M}_{\ast}\gtrsim 10^{10.5} \textit{M}_{\odot}$), where the star-forming components seem to be smaller than the corresponding stellar components. A large proportion of these galaxies have very high $Re_\text{optical}/Re_\text{MIR}$ values (i.e., larger than $1.51$, see Sect. \ref{three-classification}), suggesting the emergence, at high stellar masses, of a sizeable population of galaxies with compact star-forming cores embedded in larger stellar disks. The emergence of this population at high stellar mass naturally translates into a MIR mass--size relation with a slightly shallower slope (0.12 versus 0.17) than the one in the optical. 

In Fig. \ref{re_z}, we compare the redshift evolution of the optical and MIR sizes in three different stellar mass bins. Given that the stellar mass bins are relatively broad here, we normalized the MIR and optical sizes, accounting for the mass dependence in each stellar mass bin. To be specific, for each individual galaxy, we normalized the measured size according to the stellar mass at the center of each mass bin, while maintaining the distance of the galaxy size from the corresponding mass--size relation. 
Therefore, Fig. \ref{re_z} purely illustrates the redshift evolution of the mass--size relation. 
We find that the sizes of star-forming components of SFGs decrease by $\sim0.3$ dex from $z\sim0.1$ to $z\sim2.5$. At low and intermediate stellar masses, most SFGs have a similar size in terms of their stellar and star-forming components. In contrast, the stellar component of the most massive SFGs ($\textit{M}_{\ast}>10^{10.5}\textit{M}_{\odot}$) is around 20\% larger than the star-forming component. This results in an excess number of galaxies with a very large optical-to-MIR size ratio at high stellar mass. 

\begin{figure*}
    \centering
    \includegraphics[width=15.5cm]{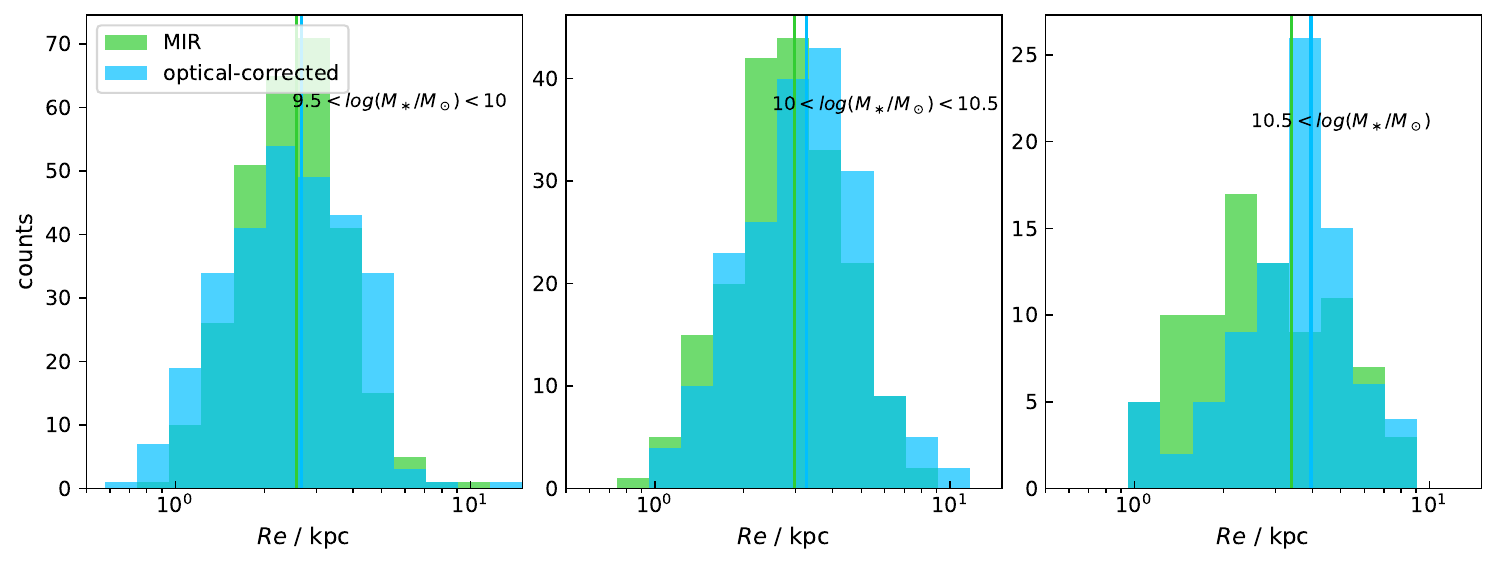}
    \caption{Distribution of optical  (\textit{blue}) and MIR (\textit{green}) effective radius, in three stellar mass bins. The vertical lines are the mass--size--redshift relation at the median redshift of 1.5 and the central stellar mass of each bin.}
    \label{re_z_dis}
\end{figure*}

Finally, in Fig. \ref{re_z_dis}, we compare the distribution of optical and MIR sizes, in three stellar mass bins, accounting for both the stellar mass and redshift dependencies of the mass--size relation. To this end, we normalized these sizes according to the stellar mass at the center of each mass bin and assuming $z\sim1.5$, while maintaining the distance of each galaxy sizes to the corresponding mass--size relations.
At low stellar masses, the two distributions nearly perfectly overlap, with both of them exhibiting slightly skewed tails toward small sizes. At intermediate masses, the peak positions of the two distributions begin to show a very small deviation: the optical size distribution is still skewed toward the small, while the MIR size distribution does not show such a skewed tail but rather exhibits a larger scatter and a center slightly shifted toward smaller sizes. 
This increase in the scatter of the MIR size distribution and its shift toward smaller sizes are even more pronounced at higher stellar masses. Overall, these size distributions can be explained by (i) SFGs consisting of two populations, one with similar stellar and star-forming sizes and the other with compact star-forming components embedded in extended stellar components; (ii) the relative number fraction of these populations changing with stellar mass; and (iii) their fractions in number being similar at the highest stellar mass. In the next section, we explore this scenario in more detail.

\subsection{Compact star-forming components}
\label{three-classification}
\begin{figure}
\begin{minipage}[c]{0.95\columnwidth}
\center{\includegraphics[width=0.99\linewidth]{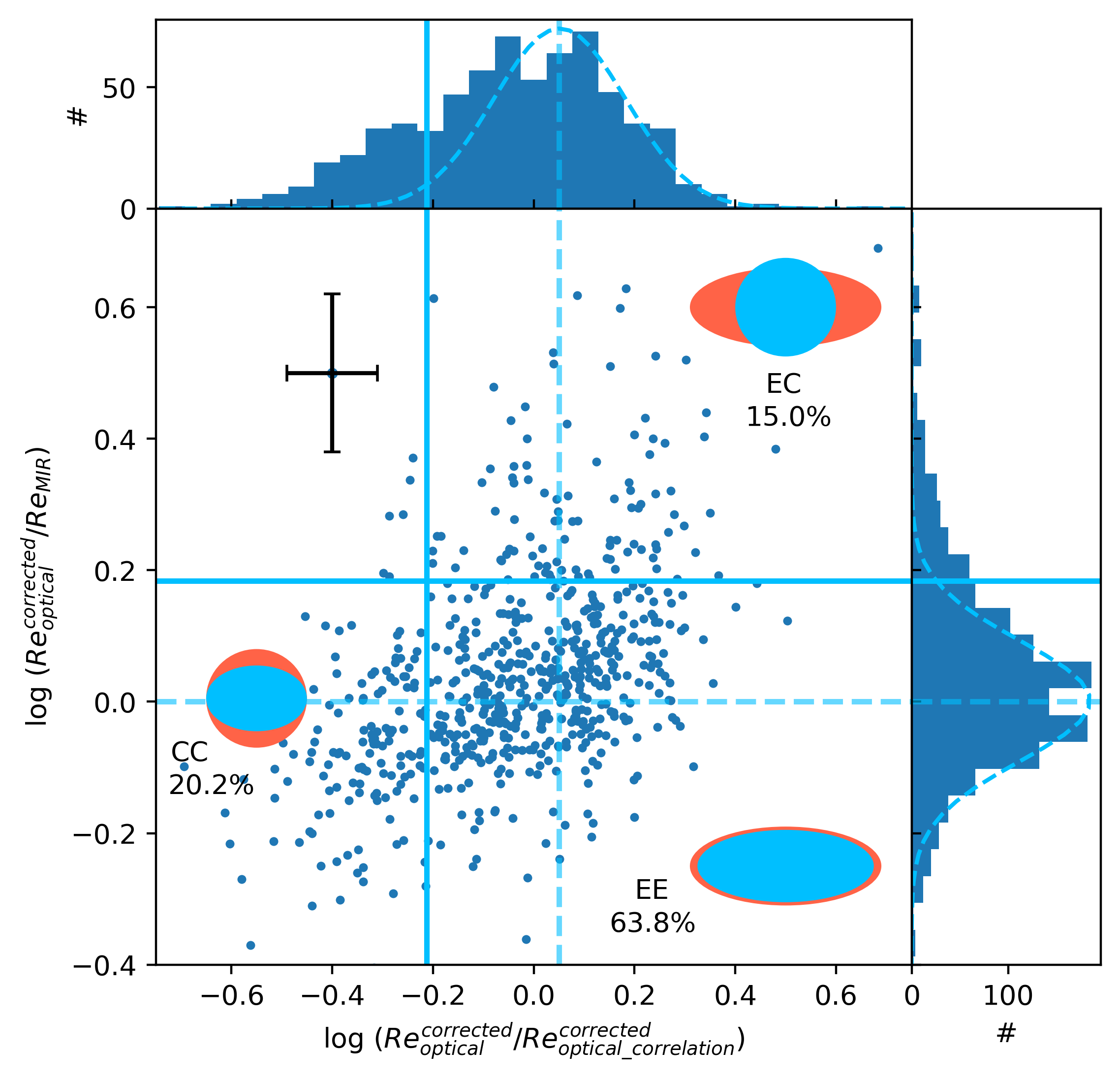}}
\end{minipage}
\caption{$\Delta Re_\text{optical}$ (distance to the optical mass--size relation) compared with the optical-to-MIR size ratio for our strategy 2 sample. The median error bar
is shown in black in the upper left. A Gaussian function is fit to both the $\Delta Re_\text{optical}$ and the size ratio distributions, as the dashed blue curve shows. Dashed blue lines show the median of the fitting, while solid blue lines show the $2\sigma$ threshold. Based on the median values for each category of galaxy in Table \ref{try1}, we sketched a schematic representation of the three galaxy classifications. Orange and blue regions are stellar and star-forming components, respectively. EE stands for extended stellar and extended SFGs. EC stands for extended stellar and compact SFGs. CC stands for compact stellar and compact SFGs. The number fraction of each category of galaxies are also given below each sketch.}
\label{classification55}
\end{figure}
The previous section shows that the majority of galaxies exhibit similar sizes in both MIR and  optical, and these galaxies tend to adhere to a well-constrained correlation between stellar mass and size. However, there are still some galaxies showing significant compact (dust-obscured) star-forming cores, especially at high stellar mass. To better comprehend these galaxies, we have to firstly distinguish which galaxies can be considered as compact or extended in terms of their stellar and star-forming components. In previous works, \citet{Barro2013} used a fixed threshold (i.e., $\text{log}(\textit{M}_{\ast}/Re^{1.5})>10.3$) to define SFGs with compact stellar components; \citet{Dokkum2015} used a slightly different criterion (i.e., $\text{log}(\textit{M}_{\ast}/Re)>10.7$) to select massive galaxies with compact stellar components, while \citet{Barro2017} derived the stellar mass surface density inside 1 kpc ($\Sigma_1^{\textit{M}_{\ast}}$), and defined SFGs that follow the $\textit{M}_{\ast}$--$\Sigma_1^{\textit{M}_{\ast}}$ relation of QGs as those with compact stellar components. In this work, following the approach in \citet{Magnelli2023}, we calculated $\Delta Re_\text{optical}=Re_{\text{individual}}/Re_{\text{fitting}}$ (i.e., the distance from the optical mass--size relation) to define whether the stellar component of a galaxy was compact or extended relative to the overall population, while taking its stellar mass and redshift dependence into account. We then calculated galaxies' optical-to-MIR size ratio to define whether their star-forming components were compact or extended relative to their stellar components. In Fig. \ref{classification55}, we compare $\Delta Re_\text{optical}$ with the optical-to-MIR size ratios. 

The $\Delta Re_\text{optical}$ and optical-to-MIR size ratio distribution reveal two things: firstly, the optical size of SFGs does not distribute as a Gaussian function but instead is skewed toward small sizes \citep[see also ][]{vdw2014}; secondly, the optical-to-MIR size ratio also exhibits a skewed distribution, revealing the emergence, at its tail, of a population of galaxies with compact star-forming components embedded in extended stellar components. To highlight these skewed distributions, we fit the $\Delta Re_\text{optical}$ and the size ratio distributions with a Gaussian function but restricting our fits to their positive and negative values, respectively. These fits allow us to identify ``outliers'' materialized by an excess of galaxies in the tail of these distributions. To separate these outliers from the general population, we simply used $-2\sigma$ and $+2\sigma$ thresholds from the center of these Gaussian distributions, respectively. 
Because the dispersions of the $\Delta Re_\text{optical}$ and size ratio distributions ($\sigma\sim$ 0.14 and 0.09 dex, respectively) are dominated by the intrinsic dispersion of the population, the definition of these thresholds are only weakly affected by the uncertainties associated with the measurements ($\sim0.03$ dex). Using these thresholds, we have effectively classified our SFGs into three main categories:
\begin{enumerate}
 \item EE galaxies (extended stellar components and extended star-forming components): $\Delta Re_\text{optical}>0.59$ and $\text{size ratio }<1.51$;
 \item EC galaxies (extended stellar components and compact star-forming components): $\Delta Re_\text{optical}>0.59$ and $\text{size ratio }>1.51$;
 \item CC galaxies (compact stellar components and compact star-forming components): $\Delta Re_\text{optical}<0.59$ and $\text{size ratio }<1.51$.
\end{enumerate}
In Fig. \ref{classification55}, we sketch the relative morphology of the stellar component (orange) and star-forming component (blue) of these three categories. We note that we speak of ``populations'' or ``categories'' when it is clear that SFGs are not separated into three different locations in this parameter space, with few galaxies in between. Rather, our definition identifies galaxies that are outliers of the general population and are in a relatively extreme phase of their evolution. The terms population or category are therefore used for the sake of simplicity. Finally, we note that although there is formally a fourth category of galaxies with compact stellar components and even more compact star-forming components, the rarity of this population (five galaxies in the strategy 2 sample; zero galaxies in the strategy 1 sample; maybe scattered out of the other three categories due to measurement errors) leads us to simply ignore it in the remainder of the analysis. In summary, approximately 66\% of the galaxies in our strategy 2 sample are classified as EE galaxies, 15\% are classified as EC galaxies, and 19\% are classified as CC galaxies. Galaxies falling into this latter category are often called blue nuggets in the literature, and are commonly thought to be the prerequisite stage before SFGs finally quench into QGs \citep{Barro2013, Dokkum2015, Tacchella2016, Barro2017, Wang2018, Lapiner2023}.

\begin{figure}
\begin{minipage}[c]{0.95\columnwidth}
\center{\includegraphics[width=0.99\linewidth]{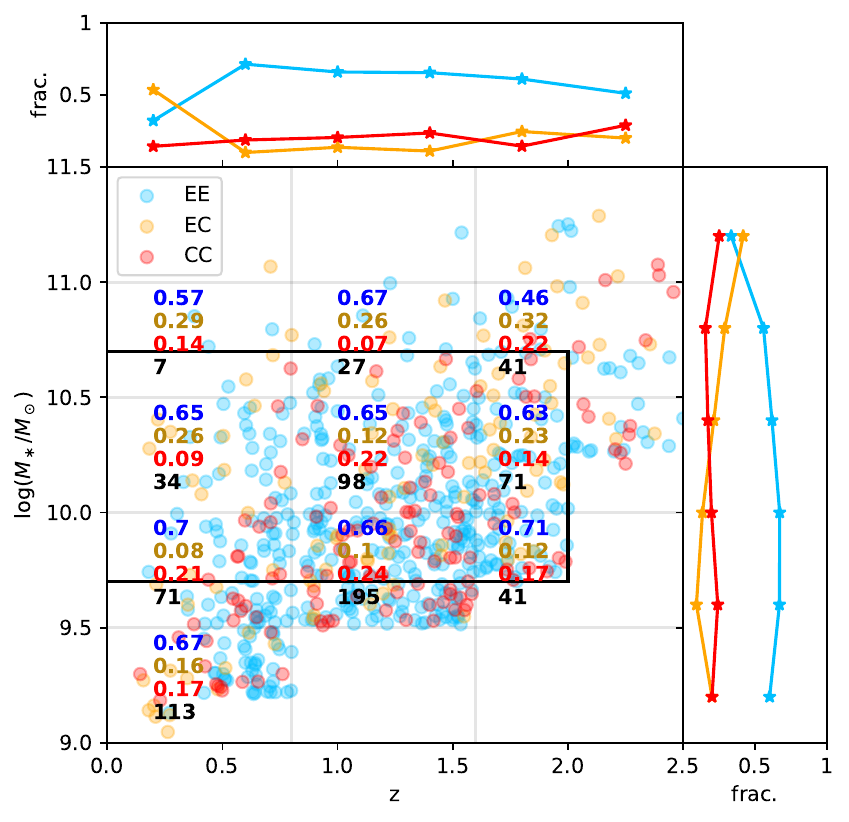}}
   \caption{Stellar mass vs. redshift for our three categories of galaxy. The top and right sidebars show the corresponding number fraction for each category of galaxy. Colored numbers are the number fraction of three categories of galaxy in the corresponding mass and redshift bins, and black numbers are the counts of galaxies in each box. The bold black box illustrates the region of the parameter space we used to build our mass-redshift-matched sample; that is, $10^{9.7}<(\textit{M}_{\ast}/\textit{M}_{\odot})<10^{10.7}$ and $z<2$.}
   \label{mass-z-cla}
   \end{minipage}
\end{figure}

In Fig. \ref{mass-z-cla}, we find that the number fraction of EE galaxies decreases smoothly with the redshift from $\sim70\%\pm4\%$ at $z\sim0.5$ to $\sim50\%\pm7\%$ at $z\sim2.5$ (excluding the first redshift bin, which shows a very limited statistic). The number fraction of EE galaxies also decreases as a function of stellar mass, but more significantly from $\sim70\%$ at $\textit{M}_{\ast}\sim10^{10}M_{\odot}$ to $\sim40\%$ at $\textit{M}_{\ast}\sim10^{11}M_{\odot}$. 
These trends are mirrored by variations of the number fraction of EC galaxies, which increases from $\sim10\%$ at  $z\sim0.5$ to $\sim20\%$ at  $z\sim2.5$ and from $\sim10\%$ at $\textit{M}_{\ast}\sim10^{10}M_{\odot}$ to almost $\sim40\%$ at $\textit{M}_{\ast}\sim10^{11}M_{\odot}$.  
In contrast, the number fraction of CC galaxies remains almost constant as a function of either redshift or stellar mass.
These results indicate that it is the stellar mass that is the main driver of the galaxy compaction phase that EC galaxies appear to be in. 
Indeed, the increased number fraction of EC galaxies at higher redshift is mostly due to a mass selection bias: the stellar mass above $10^{10.5} \textit{M}_{\odot}$ where EC galaxies appear most often is poorly represented in our low-redshift sample due to the limited comoving volume probed, whereas it dominates our sample at high redshift due to the limited stellar mass range probed.  

In Table \ref{try1}, we present the median physical properties of our three categories of galaxy. We have turned here to our strategy 1 sample to include the S{\'e}rsic index (i.e., the concentration of these galaxies) in our analysis. To account for the mass selection bias of our sample, we also calculated these median values for mass-redshift-matched samples of these three categories; that is, restricting them to the stellar-mass between $10^{9.7}$ $\textit{M}_{\odot}$ and $10^{10.7}$ $\textit{M}_{\odot}$ and redshift smaller than 2. 
For both the whole galaxy samples and the mass-redshift-matched samples, we find that (i) EE galaxies have a disk-like stellar component ($n_\text{optical}\sim1$, $b/a\sim0.5$) and a disk-like star-forming component ($n_\text{MIR}\sim0.7$, $b/a\sim0.5$); (ii) EC galaxies have a more bulgy stellar component ($n_\text{optical}>1$ but $b/a\sim0.5$) and the growth of this bulge is still going on ($n_\text{MIR}>1$, $b/a>0.5$); and finally (iii) CC galaxies have a compact, bulge-dominated stellar component ($n_\text{optical}>1$, $b/a>0.5$) and show residual growth of this bulge ($n_\text{MIR}\sim1$ but $b/a>0.5$). These results suggest that the EC and CC phases might correspond to an in situ bulge growth not seen in the EE phase and might be compatible with the wet compaction scenario advocated by simulations \citep{Tacchella2015, Lapiner2023}. This is further discussed in Sect. \ref{discussion}.

\begin{table*}
\centering  
\caption{Physical properties of galaxies in the three categories}   
\label{try1}
\tabcolsep=0.1cm
\renewcommand\arraystretch{2}
\begin{tabular}{c c c c c c c c c c c}   
\hline\hline                 
Category(\#) &$z$ &log($\textit{M}_{\ast}/\textit{M}_{\odot}$) &SFR$/\textit{M}_{\odot}\text{yr}^{-1}$ &$\Delta$MS  &$n_{\text{optical}}^{\text{corrected}}$ &$n_{\text{MIR}}$ &$Re_{\text{optical}}^{\text{corrected}}/$kpc &$Re_{\text{MIR}}/$kpc & $b/a_{\text{optical}}^{\text{corrected}}$&$b/a_{\text{MIR}}$\\   
\hline 
CC(74) & $1.19_{-0.56}^{+0.51}$&$10.04_{-0.31}^{+0.44}$&$17.11_{-11.11}^{+41.69}$&$0.16_{-0.24}^{+0.16}$&$1.35_{-0.55}^{+1.31}$&$0.77_{-0.36}^{+0.46}$&$1.66_{-0.34}^{+0.41}$&$1.8_{-0.43}^{+0.59}$&$0.74_{-0.22}^{+0.13}$&$0.78_{-0.16}^{+0.09}$\\
EE(254) & $1.15_{-0.51}^{+0.45}$&$10.1_{-0.3}^{+0.37}$&$17.28_{-11.29}^{+44.49}$&$0.11_{-0.2}^{+0.21}$&$0.91_{-0.35}^{+0.7}$&$0.64_{-0.22}^{+0.47}$&$3.73_{-1.1}^{+1.56}$&$3.5_{-0.93}^{+1.7}$&$0.51_{-0.21}^{+0.28}$&$0.56_{-0.19}^{+0.24}$\\
EC(56) & $1.38_{-0.69}^{+0.52}$&$10.35_{-0.46}^{+0.45}$&$31.56_{-23.97}^{+106.38}$&$0.2_{-0.28}^{+0.21}$&$1.42_{-0.73}^{+1.58}$&$1.23_{-0.54}^{+0.73}$&$4.04_{-1.04}^{+1.86}$&$2.13_{-0.97}^{+0.94}$&$0.49_{-0.18}^{+0.27}$&$0.64_{-0.18}^{+0.18}$\\
CC$^{\ast}$(58) & $1.2_{-0.46}^{+0.32}$&$10.05_{-0.24}^{+0.41}$&$18.17_{-8.88}^{+23.45}$&$0.16_{-0.23}^{+0.15}$&$1.47_{-0.65}^{+1.23}$&$0.85_{-0.34}^{+0.4}$&$1.66_{-0.33}^{+0.44}$&$1.85_{-0.49}^{+0.54}$&$0.76_{-0.17}^{+0.11}$&$0.8_{-0.15}^{+0.08}$\\
EE$^{\ast}$(211) & $1.18_{-0.5}^{+0.4}$&$10.09_{-0.23}^{+0.32}$&$17.39_{-9.33}^{+28.67}$&$0.11_{-0.2}^{+0.2}$&$0.88_{-0.32}^{+0.7}$&$0.64_{-0.22}^{+0.38}$&$3.73_{-1.1}^{+1.54}$&$3.59_{-0.98}^{+1.51}$&$0.5_{-0.18}^{+0.29}$&$0.57_{-0.2}^{+0.22}$\\
EC$^{\ast}$(38) & $1.24_{-0.53}^{+0.48}$&$10.24_{-0.34}^{+0.31}$&$24.78_{-14.56}^{+43.77}$&$0.1_{-0.26}^{+0.25}$&$1.41_{-0.76}^{+1.32}$&$1.25_{-0.46}^{+0.51}$&$4.21_{-1.16}^{+1.71}$&$2.23_{-0.83}^{+0.84}$&$0.48_{-0.17}^{+0.23}$&$0.59_{-0.14}^{+0.21}$\\
\hline 
\end{tabular}
\tablefoot{Median values of physical properties for galaxies in the three categories defined in Fig. \ref{classification55}. Upper and lower ranges are derived from the 16th and 84th percentiles. For those labeled with ``$\ast$,'' the sample is limited to $10^{9.7}<(\textit{M}_{\ast}/\textit{M}_{\odot})<10^{10.7}$ and $z<2$.}
\end{table*}
\renewcommand\arraystretch{1}

\subsection{Surface density}
\label{Surface density}
To go further, we calculated the stellar mass surface density within the inner 1 kpc of these galaxies ($\Sigma_{1}$), which offers a more intuitive interpretation of their concentration than to analyze their effective radius and S{\'e}rsic index separately. $\Sigma_{1}$ also provides a ``cosmic clock'' as it cannot decrease but can solely increase during a galaxy’s mass assembly \citep{Barro2017}. Finally, \citet{Barro2017} found that $\Sigma_{1}$ is a more reliable structural parameter than the surface density inside the effective radius ($\Sigma_{Re}$): the $\Sigma_{1}$--$\textit{M}_{\ast}$ relation is tighter and barely evolves with redshift compared to the $\Sigma_{Re}$--$\textit{M}_{\ast}$ relation. In this work, we adopted the equation from \citet{Ciotti1999} to analytically derive $\Sigma_{1}$:
\begin{equation}
\centering
\Sigma_{1}^{\textit{M}_{\ast}} = \textit{M}_{\ast} \times\frac{\gamma(2n_{\text{optical}}^{\text{corrected}}, b\times {Re_{\text{optical}}^{\text{corrected}}}^{-1/n_{\text{optical}}^{\text{corrected}}})}{\pi \Gamma(2n_{\text{optical}}^{\text{corrected}})},
\label{eq42}
\end{equation}
\begin{equation}
\centering
\Sigma_{1}^{\text{SFR}} = \text{SFR} \times\frac{\gamma(2n_{\text{MIR}}, b\times Re_{\text{MIR}}^{-1/n_{\text{MIR}}})}{\pi \Gamma(2n_{\text{MIR}})\pi},
\label{eq44}
\end{equation}
where $\gamma$ is the lower (left) incomplete gamma function, $\Gamma$ is the complete gamma function, and $b$ is a S{\'e}rsic index-dependent normalization parameter that satisfies
\begin{equation}
\centering
\Gamma(2n)=2\gamma(2n,b).
\label{eq45}
\end{equation}
We numerically solved Eq. \ref{eq45} using the \texttt{scipy} function \texttt{gammaincinv} to get the value of $b$. We note that $\Sigma_1$ is a projected quantity that can be affected by inclination effect. However, given that in the following we focus primarily on the median values of each population, such projection effects should not introduce significant bias into our results, assuming that the inclinations distribute uniformly across the different populations.

\begin{figure}
\center{\includegraphics[width=0.92\linewidth]{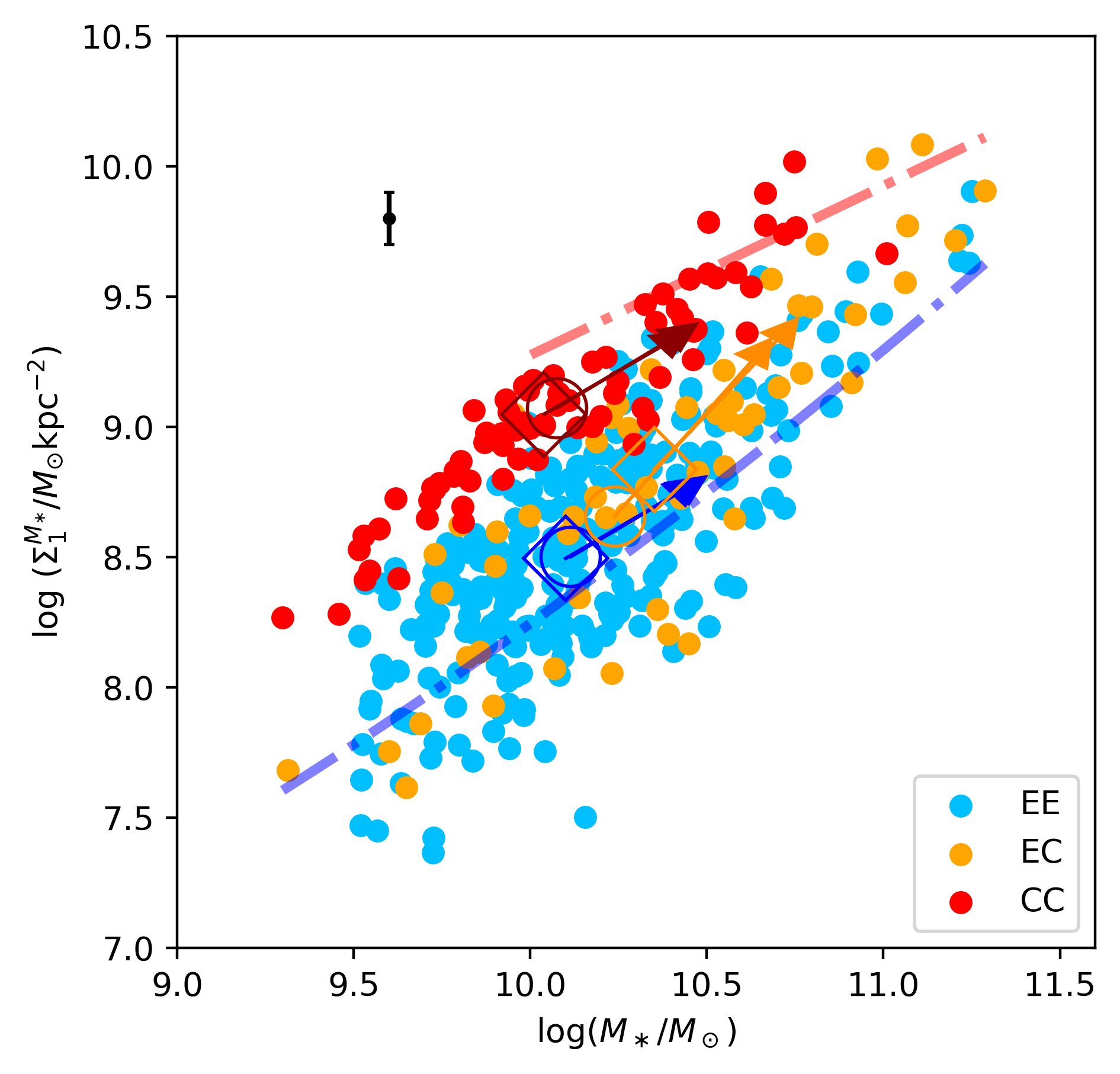}}
    \caption{Stellar mass surface density inside 1 kpc versus stellar mass. The red, orange, and blue dots represent CC, EC, and EE galaxies, respectively. The red and blue lines are $\Sigma_1^{\textit{M}_{\ast}}$--$\textit{M}_{\ast}$ relations from \citet{Barro2017} for QGs and SFGs, respectively. Diamonds represents the median value for our three categories of galaxy, and circles are the median value when restricting these galaxies to those with $10^{9.7}<(\textit{M}_{\ast}/\textit{M}_{\odot})<10^{10.7}$ and $z<2$. The arrows show how these median values would evolve in the next 600~Myr by assuming that the SFRs of these galaxies remain constant over this period and that newborn stars follow the light distribution observed in the MIR.}
\label{sigma}
\end{figure}
In Fig. \ref{sigma}, we study how our three categories of galaxy (EE, EC, and CC; see Sect. 3.5) distribute in the $\Sigma_1^{\textit{M}_{\ast}}$--$\textit{M}_{\ast}$ plane and with respect to the corresponding correlations followed by SFGs and QGs and parameterized in \citet{Barro2017}. Three main conclusions can be drawn from this comparison: firstly, for all SFGs, $\Sigma_1^{\textit{M}_{\ast}}$ is positively and tightly correlated with stellar mass, with a slope that is very consistent with the one found in \citet{Barro2017} for SFGs. As SFGs grow in their stellar mass, they build a denser central region but the sub-unity slope ($\sim0.9$) of the $\Sigma_1^{\textit{M}_{\ast}}$ relation indicates a steady inside-out structural growth of these galaxies \citep[see also][]{Barro2017}; secondly (and by construction), EE galaxies and CC galaxies follow distinct relations in the $\Sigma_1^{\textit{M}_{\ast}}$--$\textit{M}_{\ast}$ plane, with CC galaxies having at any stellar mass a higher central density than EE galaxies.
In fact, CC galaxies have structural properties similar to those of the so-called blue nuggets that were defined in \citet{Barro2017} as SFGs with stellar cores as dense as QGs (i.e., falling within 2$\sigma$ of the $\Sigma_1^{\textit{M}_{\ast}}$--$\textit{M}_{\ast}$ relation for QGs);
thirdly, EC galaxies have stellar core structures that are mainly indistinguishable from those of EE galaxies, and therefore probably share the same origin as these EE galaxies. Naturally, and by construction, EE and EC galaxies are separable on the basis of their star-forming components, and therefore their future.

Based on the galaxies' current star formation distribution, we can predict how their stellar mass distribution will evolve, and thus how they will move on the $\Sigma_1^{\textit{M}_{\ast}}$--$\textit{M}_{\ast}$ plane. To this end, for each individual galaxy, we have made two simplified assumptions: that the current SFR remains constant for the next 600 million years \citep[i.e., the average cold gas depletion time observed in these SFGs; ][]{Wang2022} and that all newly born stars are distributed according to the MIR light profile. Therefore, the newly added stellar mass is SFR $\times$ 600 Myr and the future stellar component is distributed according to the sum of its current optical and MIR light profiles. In Fig. \ref{sigma}, we overplot these predictions in the form of arrows. 
We find that EE galaxies will grow along the scaling relation of SFGs, and thus will not go through a size compaction phase that is necessary to explain the formation of CC galaxies.
In contrast, the compact star-forming component in EC galaxies will naturally result in a tremendous increase in their $\Sigma_1^{\textit{M}_{\ast}}$, bringing them into the sequence of CC galaxies. 
Finally, CC galaxies, with their already compact stellar cores, will secularly evolve toward the scaling relation followed by QGs. 

All these results support an evolutionary path from EE galaxies to EC galaxies, then to CC galaxies and, finally, to QGs. Our unique view of the star-forming components of galaxies has therefore enabled us to identify a population of galaxies passing through this critical phase of compaction, which is a prerequisite for the formation of QGs. Studying the mechanism behind EC galaxies is now crucial to our understanding of galaxy evolution.

\subsection{Caveats}
There are two main caveats that may affect the use of MIR sizes as a proxy for the sizes of the star-forming components in galaxies. One caveat is the representativeness of dust-obscured SFR (i.e., ${\rm SFR}_{\rm IR}$) compared to the total SFR (i.e., ${\rm SFR}_{\rm UV}$+${\rm SFR}_{\rm IR}$). By comparing the dust-unobscured UV SFRs and dust-obscured SFRs of our SFGs deduced from the CIGALE SED fits, we find that the former are subdominant with the ratio of ${\rm SFR}_{\rm UV} / ({\rm SFR}_{\rm IR}+{\rm SFR}_{\rm UV}$) well below 30\%. These results are consistent with the literature for the same stellar mass range probed here \citep[$\textit{M}_{\ast}>10^{9.5}\textit{M}_{\odot}$;][]{Whitaker2012, Shen2023, Magnelli2023, Shivaei2024}. Therefore, we consider that, on global scales, the dust-obscured star-forming components measured by MIRI trace the bulk of the star formation activities in our galaxies. 
However, ${\rm SFR}_{\rm UV} / ({\rm SFR}_{\rm IR}+{\rm SFR}_{\rm UV})$ might change within the galaxies (i.e., dust attenuation gradient). Unfortunately, not much is known about such variation, but one can assume that the inner parts of galaxies are typically more dust-obscured than their outskirts \citep{Miller2022}. This makes ${\rm SFR}_{\rm IR}$ more representative in the galaxy center than in the outskirts, and hence decreases the observed MIR half-light radius compared with the intrinsic half-SFR radius. Quantifying the magnitude of this effect is beyond the scope of our study, and we defer this analysis to a future pixel-scale study, combining JWST/NIRCam and JWST/MIRI observations. Nevertheless, it should be borne in mind that the number fraction of compact SFGs in our study may be slightly overestimated due to this effect.

Another caveat is the representativeness of MIR emission compared to the total infrared emission \citep[often expressed by the index IR8 $ \equiv L_\text{IR}/L_{8}$;][]{Elbaz2011}. It has been found that on global scales and for main-sequence galaxies, IR8 equals $\sim4$ and is independent of redshift and stellar mass \citep{Schreiber2018}. This implies that for most galaxies our MIRI 18 $\mu$m observation (rest-frame about 8 $\mu$m) is a good proxy for ${\rm SFR}_{\rm IR}$. 
However, it has been found that, on global scales, IR8 increases when galaxies are above the main sequence (i.e., for starbursts) or when galaxies have low metallicity, and hence low stellar mass \citep{Engelbracht2005, Elbaz2011, Schreiber2018, Whitcomb2020}. This change in IR8 from one galaxy to another does not influence our size measurements, as long as IR8 remains constant inside each galaxy. However, it could be problematic if IR8 varies inside a given galaxy. One can make the assumption that if such variation occurs, IR8 goes from a main-sequence-like value on the outskirts to a starburst-like value in the center. In this case, our MIR half-light radius would be larger than the intrinsic half-SFR radius of these galaxies. 
This might decrease the number fraction of compact galaxies in our study. The possible overestimation of the star-forming size of galaxies due to this variation in IR8 and the possible underestimation of the star-forming size due to the variation in ${\rm SFR}_{\rm UV} / ({\rm SFR}_{\rm IR}+{\rm SFR}_{\rm UV})$ could cancel each other out. However, we defer this conclusion to a future pixel-scale analysis.

\section{Discussion}
\label{discussion}
We analyzed the optical and MIR structural properties of SFGs in the CANDELS $\times$ MIRI COSMOS and UDS fields to study both their stellar and (dust-obscured) star-forming components. We characterized these structural properties as a function of various physical properties of these SFGs (e.g. reshift and stellar mass). We find that the stellar and star-forming components of most SFGs have disk-like structures with relatively similar sizes (i.e., EE galaxies; $66\%$), which grow linearly with their stellar mass (the so-called mass--size relation). However, there is a group of galaxies ($\sim15\%$) with compact star-forming components embedded in extended stellar components (i.e., EC galaxies). These EC galaxies are relatively rare at low stellar mass ($\sim10\%$), but this population becomes important at high stellar mass ($\textit{M}_{\ast}>10^{10.5} \textit{M}_{\odot}$), where it corresponds to $\sim30\%$ of the total SFG population. Our study suggests that these EC galaxies could be the intermediate stage between EE galaxies and a third category of SFGs with compact stellar and star-forming components (i.e., CC galaxies; $19\%$; sometime also called blue nuggets), these latter galaxies probably being the progenitors of the QGs that we observe at low redshift \citep{Barro2017}.

This evolutionary path (EE$\rightarrow$EC$\rightarrow$CC galaxies), in which the EC phase corresponds to the in situ formation of a compact bulge, could be compatible with the wet compaction scenario advocated by simulations \citep{Tacchella2015, Barro2017, Lapiner2023}. In this scenario, which implies a loss of angular momentum of the dark matter halo, a typical gas-rich galaxy with its extended disk-like structures (large size, low S{\'e}rsic index, and $\sim0.5$ axis ratio for the stellar and star-forming components; EE phase) undergoes violent disk instabilities (VDIs) that compress its gas components, triggering intense star formation, effectively transforming its star-forming component into a compact, concentrated one (small size, high S{\'e}rsic index, and high axis ratio of the star-forming components; EC phase); Finally, this intense star formation creates a dominant compact and concentrated stellar component with a structure that looks like a bulge (small size, high S{\'e}rsic index, and high axis ratio of the stellar components), but rapidly depletes the cold gas and leads to less intense star formation (small size, low S{\'e}rsic index, and high axis ratio of the star-forming component; CC phase). Therefore, although caution must be exercised due to potential complex progenitor biases, the physical properties of the EE, EC, and CC galaxies are consistent with them representing the three phases of the wet compaction scenario.

\subsection{The incidence of EC galaxies}
Our results reveal that at high stellar mass ($\textit{M}_{\ast}>10^{10.5} \textit{M}_{\odot}$), there is an emergence of a significant population of galaxies ($\sim30\%$) with compact star-forming cores embedded in the extended stellar components (i.e., EC galaxies).
This number fraction of EC galaxies is consistent with that found in \citet[][i.e., $27\%$ of their galaxies with $\textit{M}_{\ast}\geq10^{10.5} \textit{M}_{\odot}$]{Magnelli2023}.
It is also in line with the results obtained by \citet{{Puglisi2021}} in the submillimeter using ALMA. Indeed, about 40\% of their SFGs at $1.1 < z < 1.7$ and with $\textit{M}_{\ast}>10^{10.5} \textit{M}_{\odot}$ have a compact star-forming core embedded in a more extended stellar component.
The fact that EC galaxies appear predominantly at high stellar mass is also consistent with the broken luminosity (mass)--size relation described by \citet{pozzi2024}, who used ALMA to measure this relation and found a downward trend after a point at $\textit{L}_{\rm IR}>10^{12}\textit{L}_\odot$.
Finally, ALMA observations of galaxies at even higher redshift than our study (i.e., $z\gtrsim2.5$) reveal that the majority of massive ($\textit{M}_{\ast}\gtrsim10^{11}\textit{M}_{\odot}$) SFGs, if not all of them, have a compact dust-obscured star-forming component embedded in a more extended stellar component \citep{Simpson2015,Hodge2016,Fujimoto2017,CG2018,Elbaz2018,Lang2019,Puglisi2019,Gullberg2019,Chang2020,Tadaki2020,CG2022}.
Therefore, if at a given stellar mass the number fraction of EC galaxies does not increase significantly from $z\sim0$ to $z\sim2.5$, it does so at an even higher redshift. Overall, these results indicate that as SFGs grow in mass, their star-forming components are more likely to go through a compaction phase, and this probability increases further at $z>2$.

Although the existence of a population of EC galaxies undergoing a critical phase of compaction seems relatively well corroborated in the literature, the exact physical mechanism that triggers this phase remains controversial. In principle, galaxy compaction must be ``wet,'' meaning that the process has to involve a dramatic loss of angular momentum \citep{Dekel2014, Zolotov2015}. 
In the narrative of wet compaction, a gas-rich SFG can develop VDIs, which are associated with (i) major or minor wet mergers \citep{Dekel2006, Covington2011}; (ii) counter-rotating cold gas streams \citep{Danovich2015}; (iii) recycling of stellar winds or fountains \citep{Elmegreen2014}; or (iv) satellite tidal compression \citep{Renaud2014}. 
All these simulations typically predict that the occurrence of this violent compaction phase increases with the stellar mass (with low-mass galaxies going through a much milder compaction phase) and at higher redshifts where galaxies are much richer in cold gas and their denser cosmic environment leads to more frequent major or minor mergers \citep{Zolotov2015, Tacchella2015, Lapiner2023}. Although our results strongly support an increase in the occurrence of this compaction phase with the stellar mass, they only support a strong redshift evolution at $z>2$.

If there is an evolutionary path from EE to EC, and to CC galaxies, the emergence of EC galaxies at high stellar mass should also be accompanied by a rise in the number fraction of CC galaxies. Such an increase is, however, not observed, with their number fraction even slightly decreasing with stellar mass. Part of this discrepancy can be attributed to our classification strategy, in which we did not take into account the change in the dispersion of the optical mass--size relation. Indeed, the optical size distribution is slightly more dispersed at low stellar mass than at high stellar mass \citep{vdw2014}. As a result, low-mass galaxies are more likely to be classified as optically compact in our approach. After taking this into account, the number fraction of galaxies classified as CC at low stellar mass would decrease; hence, the number fraction of CC galaxies would probably increase with stellar mass. 
Nevertheless, this would not alter the fact that at high stellar masses ($\textit{M}_\ast>10^{10.5}\textit{M}_\odot$), the number fraction of CC galaxies would still be lower than that of EC galaxies ($\sim15\%$ vs. $\sim40\%$). 
This implies that the timescale of compaction is longer than that of galaxy quenching (CC galaxies becoming QGs) or galaxy rejuvenation (CC galaxies falling back into the EE category due to rejuvenation from gas inflow). Simulations only marginally support this result as they typically find comparable timescales of a few hundred million years for these three critical phases \citep{Zolotov2015, Tacchella2015, Lapiner2023}. To further investigate these timescales, more detailed studies of the gas components of EE, EC, and CC galaxies with ALMA are needed. 

\subsection{Structural evolution of star-forming galaxies}
\begin{figure}
 \center{\includegraphics[width=0.96\linewidth]{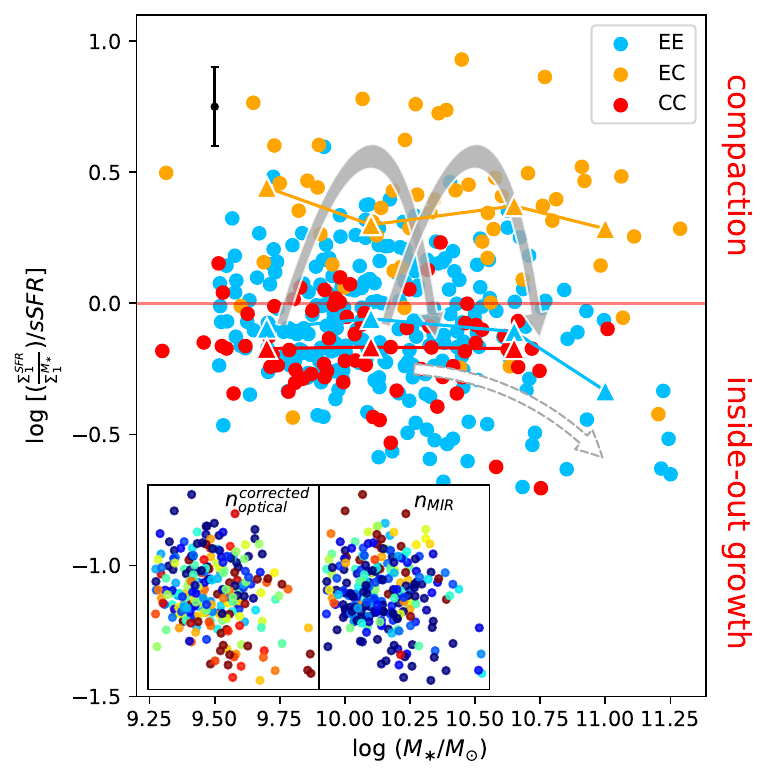}}
    \caption{Stellar mass against the ratio of the sSFR measured inside 1 kpc and that measured over the whole galaxy. Red, orange, and blue circles are CC, EC, and EE galaxies, respectively. Median values are labeled by a small triangle. The solid gray arrows depict the scenario in which the secular growth of SFG is interrupted by violent phases of compaction. The dashed arrow illustrates the inside–out quenching for massive SFGs. The two insets show the same distribution, only for EE galaxies color-coded with their optical (\textit{left}) and MIR (\textit{right}) S{\'e}rsic index (blue: low value; red: high value).}
\label{sSFR}
\end{figure}

To better illustrate the overall picture of structural evolution of SFGs, in Fig. \ref{sSFR}, we compare their stellar mass against their $\Sigma_1^{\text{sSFR}}/$sSFR, where sSFR is the ratio between SFR and stellar mass, and $\Sigma_1^{\text{sSFR}}=\Sigma_1^{\text{SFR}}/\Sigma_1^{\textit{M}_{\ast}}$. $\Sigma_1^{\text{sSFR}}/$sSFR characterizes how the inner structure of SFGs evolves with respect to their outskirts. If $\Sigma_1^{\text{sSFR}}/$sSFR $= 1$, the structure of the stellar component is not going to be changed by the ongoing star formation activity. If $\Sigma_1^{\text{sSFR}}/$sSFR $> 1$, the inner 1 kpc of the stellar component evolves faster than the outskirt (i.e., compaction). In contrast, if $\Sigma_1^{\text{sSFR}}/$sSFR $< 1$, the galaxy outskirts grow faster than its inner regions (i.e., inside-out growth). 

EE galaxies have a median value of $\Sigma_1^{\text{sSFR}}/$sSFR$\sim0.8$, indicating that this population of galaxies follows a secular inside-out growth, leading to the mass--size relation seen in Fig. \ref{mass_size_relation}. However, for some reasons (e.g., VDI), some of these EE galaxies have their star-forming components compressed, and turn into EC galaxies, exhibiting high $\Sigma_1^{\text{sSFR}}/$sSFR (i.e., $\sim2$). Subsequently, these galaxies grow a compact stellar core, leading to CC galaxies with less extreme star formation and less extreme $\Sigma_1^{\text{sSFR}}/$sSFR (i.e., $\sim0.7$). In Fig. \ref{sSFR}, such a compaction evolutionary path is sketched as an inverse-U path.

Despite the existence of this possible evolutionary scenario, there is a significant fraction of very massive ($\textit{M}_\ast>10^{10.7}\textit{M}_\odot$) EE galaxies that have apparently not yet passed through such a compaction phase. While by definition these very massive EE galaxies have extended stellar and star-forming components, they also have a fairly concentrated stellar profile (i.e., high optical S{\'e}rsic index; see inset of Fig. \ref{sSFR}), suggesting the existence of a stellar bulge in their inner cores. As was already mentioned, because in EE galaxies the star-forming components remain predominantly in a disk-like structure at all stellar masses (i.e., $n_\text{MIR}\sim0.7$; see inset in Fig. \ref{sSFR}), the presence of these massive stellar bulges is at odds with a scenario in which very massive EE galaxies grow only via this EE phase.
The existence of massive bulges inside these very massive EE galaxies suggests two different scenarios: (i) the bulge of these galaxies grows ex situ via multiple relatively dry minor and major mergers, without experiencing any compaction phase; or (ii) the bulge of these galaxies grows in situ during several violent phases of compaction (EE$\rightarrow$EC$\rightarrow$CC), which were followed by a phase of rejuvenation through the growth of a new gas ring from gas infalling or minor mergers. This rejuvenation on the galaxy outskirts is supported by numerous hydrodynamic simulations \citep{Zolotov2015, Tacchella2016, Lapiner2023} and would correspond observationally to galaxies with more extended star formation than stellar components, as is shown in some of our galaxies ($Re_\text{optical}/Re_\text{MIR} < 10^{-0.18}$; Fig. \ref{classification55}). In this case, the EC phase would naturally correspond to the in situ growth of the bulge of these very massive EE galaxies, as is suggested by the high S{\'e}rsic index and the spheroid structure (axis ratio larger than 0.5) of the star-forming components in EC galaxies.

As SFGs might go through one or more phases of compaction, their stellar cores increasingly resemble those of QGs in terms of surface density ($\Sigma_1$ of CC galaxies or $\Sigma_1$ of very massive EE galaxies; see Fig. \ref{sigma}), S{\'e}rsic index ($n_\text{optical}>1$) and structure (axis ratio larger than 0.5). If star formation ceases in these galaxies, they may turn into QGs. How star formation ceases in these galaxies is, however, still a matter of debate. Some invoke morphological quenching \citep[e.g.,][]{Martig2009}, because once a galaxy becomes bulge-dominated, its gaseous disk is stabilized against fragmenting into star-forming clumps. This quenching process operates from the inside-out, and often occurs at high stellar mass when galaxies have had time to build massive bulges. Others invoke the suppression of gas inflow due to various heating mechanisms taking place in very massive halos \citep[e.g.,][]{Tacchella2015}, followed by a high (virial) temperature being maintained due to radio AGN feedback \citep[e.g.,][]{Chen2019}. Finally, others invoke the gas removal by AGN feedback, which is only effective in galaxies with sufficiently massive supermassive black holes \citep[e.g.,][]{Weinberger2018}. Although it is impossible with our observations to distinguish between these scenarios, we have clues that this quenching operates from the inside-out, especially for very massive EE galaxies. Indeed, in Fig. \ref{sSFR}, we observe an anticorrelation between stellar mass and $\Sigma_1^{\text{sSFR}}/$sSFR for very massive EE galaxies, with a Pearson correlation of $-0.2$. This favors the scenario of morphological quenching or AGN quenching over that of gas supply suppression, which typically does not operate from the inside-out. 

It should be noted that the galaxy evolution scenario presented here is based on the simplified assumption that increasing stellar mass or increasing stellar mass density is equivalent to a cosmic clock. However, the real scenario of galaxy evolution might be more complex. For example, some galaxies can undergo large jumps in stellar mass due to bursty star-formation activities, while some other galaxies can increase their stellar mass through minor and major mergers. These lead to breaks in their relative cosmic clock. Overall, care must be taken when using snapshots of galaxies in a limited range of redshifts to sketch out an evolutionary model, and more sophisticated analyses of their SFHs are needed.

\section{Summary}
\label{summary}
In this paper, we have used the imaging from PRIMER JWST/MIRI \textit{F1800W} to study the MIR morphology of SFGs in the CANDELS COSMOS and UDS fields. The rest-frame MIR emission allows us to study the dust-obscured star-forming components in SFGs. Our final mass-complete sample contains 384 (strategy 1) and 665 (strategy 2) SFGs at $0<z<2.5$, with $M_\ast\gtrsim10^{9.5}\textit{M}_{\odot}$ and $\Delta MS>-0.5$. We crossmatched our MIRI detected sample with the rest-frame optical observations from CANDELS, to retrieve the associated physical parameters (e.g., redshift, stellar mass, rest-frame optical morphology, etc.). We compared the rest-frame MIR and optical morphologies of these galaxies to study the intrinsic relation between their star-forming and stellar components, thereby exploring the structural evolution of SFGs. Our main results can be summarized as follows:
\begin{enumerate}
 \item 
 The stellar and star-forming components of SFGs have a median S{\'e}rsic index of $1.05^{+0.93}_{-0.50}$ and $0.69^{+0.66}_{-0.26}$, respectively, and their axis ratio distributions are relatively flat with a median value of $\sim0.5$. This implies that the stellar and star-forming components in most SFGs have disk-like structures, which are actually well aligned, as is indicated by their small astrometric offset and consistent position angle. The slightly lower S{\'e}rsic index of the star-forming components might, however, indicate the presence of more clumpy structures in the star-forming disk than in the stellar disk.
 \item Despite peaking at $n\sim1$, the S{\'e}rsic index of the stellar component increases with stellar mass, implying that as SFGs grow in mass, their stellar components become more centrally concentrated, likely because of an evermore dominating central bulge. Because the star-forming components remain mostly in a disk-like structure at all stellar masses, these bulges do not seem to form in situ in a secular process.
 \item The sizes (effective radius) of the stellar and star-forming components of SFGs are generally the same. Consequently, the stellar and star-forming components follow a very similar mass--size--redshift relation in which the size increases as SFGs become more massive or are observed at a lower redshift.
 \item The size distribution for both the stellar and star-forming components is skewed toward small values. By comparing those sizes with the average values, we classified SFGs into three categories: EE galaxies (extended stellar components and extended
 star-forming components, 66\%), EC galaxies (extended stellar components and compact
 star-forming components, 15\%), and CC galaxies (compact stellar components and compact
 star-forming components, 19\%). The number fraction of EC galaxies increases from 10\% at low stellar mass ($\textit{M}_{\ast}\sim10^{10}\textit{M}_{\odot}$) to 30\% at high stellar mass ($\textit{M}_{\ast}>10^{10.5}\textit{M}_{\odot}$), mirrored by the decrease in the number fraction of EE galaxies over the same stellar mass range. Stellar mass thus seems the main driver of the emergence of this EC population.
 \item The inner 1 kpc stellar mass surface density ($\Sigma_1^{\textit{M}_{\ast}}$) of EE and EC galaxies are very similar, indicating that these two populations might share the same origin. In contrast, CC galaxies have a value $\sim0.5$ dex higher than EE and EC galaxies, close to the $\Sigma_1^{\textit{M}_{\ast}}$ of QGs. The inner 1 kpc SFR surface density ($\Sigma_1^\text{SFR}$) of EE galaxies has a lower value than that of EC galaxies, suggesting that these latter galaxies are in a phase of violent out-of-equilibrium compaction, building dense stellar bulges in situ ($n_\text{MIR}>1$ and $b/a_\text{MIR}>0.5$). This compaction phase is significant enough to increase the stellar mass surface density of EE galaxies to the very large value observed in CC galaxies. This points to an evolutionary path from EE to EC to CC galaxies.
\end{enumerate}
In summary, our results indicate that the structural evolution of the stellar component of SFGs is initially due to an inside-out secular growth (EE galaxies), which leads to the establishment of the optical mass--size relation. However, this secular growth can be interrupted by one or more violent phases of compaction (EC galaxies), triggered by internal or external mechanisms, which build in situ stellar cores resembling those of QGs (CC galaxies or very massive EE galaxies). It remains unclear whether this phase of compaction is unique and followed by quenching, or whether it is followed by a phase of rejuvenation that brings the galaxies back into the EE population. The latter scenario is favored by the existence in massive EE galaxies of bulges that seem to form only during this EC phase, unless ex situ bulge formation plays an important role.
To better understand the structural evolution of SFGs, and in particular the mechanism leading to this critical compaction phase in which EC galaxies find themselves, high-resolution observations of their gas reservoir with ALMA will be necessary.

\begin{acknowledgements}
YL, BM, DE, and CGG acknowledges support from CNES. This work was supported by the Programme National Cosmology et Galaxies (PNCG) of CNRS/INSU with INP and IN2P3, co-funded by CEA and CNES. YL acknowledges the following open source software used in the analysis: Astropy \citep{Astropy2022}, photutils \citep{photutils}, PetroFit \citep{PetroFit} and NumPy \citep{numpy}. PGP-G acknowledges support from grant PID2022-139567NB-I00 funded by Spanish Ministerio de Ciencia e Innovaci\'on MCIN/AEI/10.13039/501100011033,
FEDER \textit{Una manera de hacer Europa}. DJM acknowledges the support of the Science and Technology Facilities Council.
\end{acknowledgements}

%
\bibliographystyle{aa.bst} 
\bibliography{BiB.bib} 
%

\begin{appendix}
\section{} 
In this appendix we show the scatter plot of stellar mass, redshift and the ratio between the optical and MIR effective radius of the SFGs in our sample, as a supplementary figure to Fig.\ref{re_change}.
\begin{figure}[h]
\centering
\includegraphics[width=0.9\columnwidth]{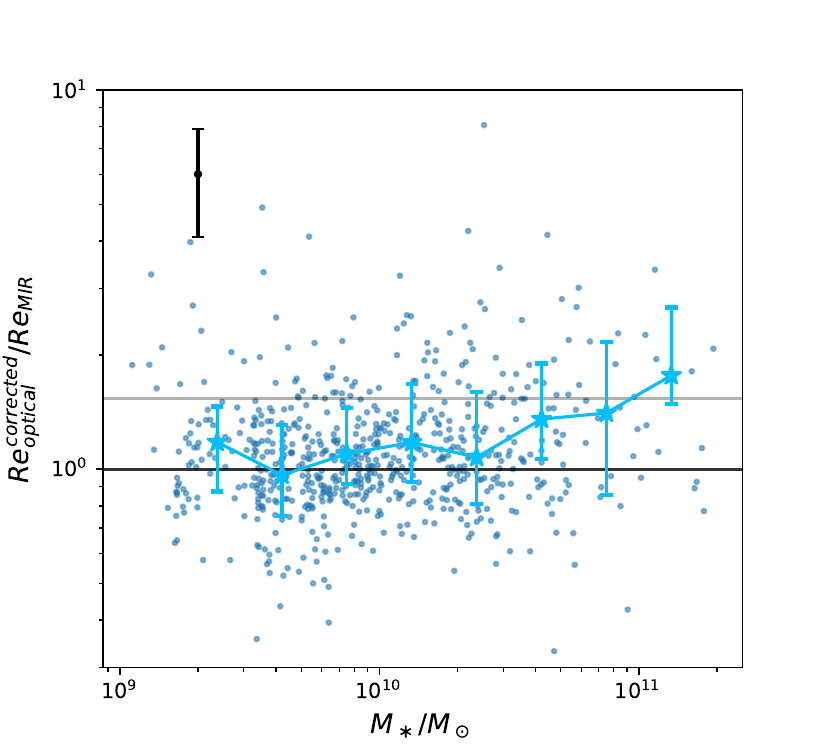}
\includegraphics[width=0.9\columnwidth]{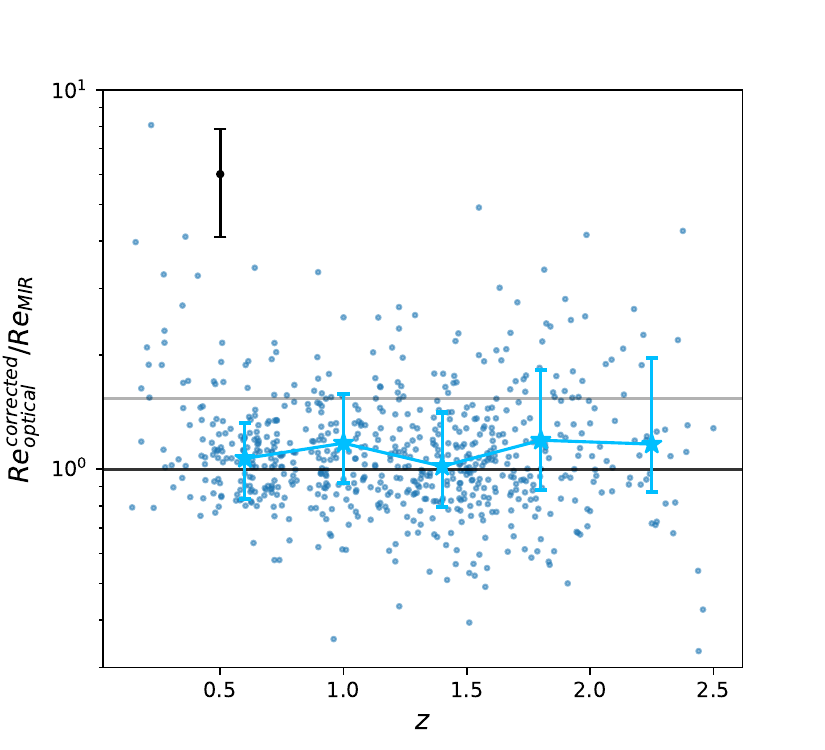}
\caption{Comparison of stellar mass (\textit{top}), redshift (\textit{bottom}) and the ratio between the optical and MIR effective radius of the SFGs in our sample. The black horizontal line shows the ratio of unity and the gray line shows the $2\sigma$ threshold as indicated in Fig.\ref{classification55}. Blue stars show the corresponding median value.}
\label{appendix12}
\end{figure}
\end{appendix}
\end{document}